\documentclass{article}

\usepackage{amssymb,amsfonts,amsmath,stmaryrd}
\usepackage{enumerate,float,indentfirst}
\usepackage{color}
\usepackage{diagbox}
\usepackage{tikz}
\usepackage{ytableau}
\usepackage{setspace}
\usetikzlibrary{arrows,snakes,backgrounds}
\usepackage[hidelinks,colorlinks=true,unicode]{hyperref}
\hypersetup{linkcolor=blue, citecolor=blue, filecolor=blue, urlcolor=blue}
\usepackage{url}
\usepackage[natbib=true, style=numeric-comp, sorting=none]{biblatex}
\def\be{\begin{eqnarray}}
\def\ee{\end{eqnarray}}
\def\nn{\nonumber}

\def\Tr{{\rm Tr}\,}
\def\bR{\bar R}

\definecolor{red}{rgb}{1,0,0}
\definecolor{orange}{rgb}{1,0.5,0}
\definecolor{violet}{rgb}{0.7,0,1}

\hoffset= - 1.0in         % switch off for draft style
\begin{document}

\hfill MIPT/TH-05/22

\hfill IITP/TH-06/22

\hfill ITEP/TH-08/22

\bigskip

\centerline{\Large{{\bf
Evolution properties of the knot's defect}
}}

\bigskip

\centerline{ {\bf A.Morozov, N. Tselousov} }

\bigskip

\centerline{\it MIPT, ITEP(KKTEP) \& IITP, Moscow, Russia}

\bigskip

\centerline{ABSTRACT}

\bigskip

{\footnotesize
The defect of differential (cyclotomic) expansion for colored HOMFLY-PT polynomials
is conjectured to be invariant under any antiparallel evolution
and change linearly with the evolution in any parallel direction.
In other words, each ${\cal R}$-matrix can be substituted by
an entire 2-strand braid in two different ways:
the defect remains intact when
the braid is antiparallel and changes by half of the added length
when the braid is parallel.
}

\bigskip

\bigskip

\section{Introduction}

Knot theory in $3d$ is currently among the most important topics in theoretical physics.
It is closely related to the well studied conformal Wess-Zumino-Witten model in $2d$,
which is fully controlled by group theory.
On the other hand, this is the theory where observables are the true Wilson lines,
like in $4d$ gauge theories, which we understand much worse.
Still, the theory is topological, and the averages do not depend on the shape of
integration contours, only on the way they are knotted and linked.
This puts knot theory at the border between group theory and quantum field theory,
and it is usually at such borders that the new knowledge is generated intensively and fast.
From the point of view of field theory the most important is the possibility
to go beyond perturbation theory and study the implication of symmetries for
non-pertirbative correlators.
As already known from conformal field theory in $2d$ the symmetries are naturally deformed,
and the underground machinery is that of quantum groups and ${\cal R}$-matrices.

Wilson line averages for $\mathfrak{sl}_N$ in knot theory are named HOMFLY-PT polynomials
and they are described by matrix elements of ${\cal R}$-matrix products
along various braids.
Somewhat surprisingly they are indeed {\it polynomials} in non-perturbative variables
$q=\exp\left(\frac{2\pi i}{g+N}\right)$ and $A=q^N$ (these are the standard notations
for $\mathfrak{sl}_N$, the situation for other series of simple Lie algebras is similar).
Generic representation theory constrains the dependence of HOMFLY-PT polynomials
on representation $R$, which is used to define the Wilson line
\be
H_R^{\cal K} = \left<\Tr\!_R \, {\rm Pexp}\left(\oint_{\cal K} {\cal A}\right)\right>
\ee
and these restrictions can be formulated in terms of the
{\it differential (cyclotomic) expansion} \cite{IMMM,MMM2,AMM, MMM1, BM1,Morozov1, Morozov2,Morozov3,Morozov4, Morozov5, Morozov6, BJLMMMS, KM1,KM2,KM3, Itoyama:2012qt,Itoyama:2012re,Habiro2007,Nawata:2015wya,KNTZ1,CLZ,Chen:2014fpa,Chen:2015rid,Chen:2015sol,Kawagoe:2012bt,Kawagoe:2021onh,BG, Gorsky:2013jxa,Gukov:2011ry,Dunfield:2005si, berest2021, lovejoy2019colored, lovejoy2017colored, hikami2015torus,garoufalidis2005analytic,garoufalidis2011asymptotics}.
Despite the simplicity of definitions, this expansion is rather difficult to get
and study even in particular examples, and we are still at the stage when new examples
provide new insights. For recent developments in another direction - perturbative expansion of HOMFLY-PT polynomials, see \cite{Lanina:2021jzd,Lanina:2021nfj}.

For every particular knot the differential expansion appears {\it enhanced}
over requirements of the pure representation theory,
and the level of enhancement is characterized by the {\it defect}.
For zero defect the structure is almost two times stronger than
anticipated, the enhancement decreases for larger defects, but it is always present.
The puzzling story of defects is not yet broadly known,
and there is a lot to study and understand.

The task of this paper is to 
%summarize the new ideas, inspired by extending
%the previous knowledge from restricted
%(twist and double-braid) to generic triple pretzels.
%From twist knots we learned the evolution \cite{evo} and $C$-polynomials \cite{Cpols},
%from double braids -- the important exclusive???   Racah matrices $S$ and $\bar S$ \cite{Mor16}.
%The triple pretzels generalize them, preserving the defect-zero property,
%and, dealing with such a big family, we manage to formulate a new hypothesis:
to formulate a new hypothesis: 
that {\bf defect is invariant under the antiparallel evolution and linearly changes along the parallel one}.
We test this hypothesis on various examples.
Dedicated technique for {\it proving} this kind of properties remains to be developed.

In sec.\ref{diffexp} we remind the original definition of the differential expansion (DE) from \cite{AMM,MMM1}.
Sec.\ref{defects} is devoted to the concept of {\it defect of DE}, introduced in \cite{KM1},
and surveys various hypothesis about it, old and new.
Sec.\ref{basexa} provides the two basic examples, of antiparallel triple pretzel
(which include twist and double braids) and 2-strand torus knots.
The second case illustrates the subtleties in the formulation of our conjecture
for parallel evolution.
Another kind of subtleties are introduced in sec.\ref{3pretAlex}, they concern
the "simplified" description of defect,
which is often, but not always, related to degree of the fundamental Alexander polynomial.
Calculations with this definition are tremendously easier, but additional care is needed
in interpreting the results.
%In sec.\ref{arbor}
In the Appendix at the end of the paper
we remind the basics of arborescent calculus and its generalization to a wider
class of knots -- this is what we need to consider more complicated examples in sec.\ref{compexa}.
%
%-- unfortunately there are exceptions from this simple and would-be-convenient rule,
%???and we demonstrate the way they occur and the structure they have.
%secs.4 and 5??? provide relevant examples, respectively of antiparallel and parallel
%evolutions, which explain the details of our conjecture.
%???sec.6 is devoted to the subtleties of
%%for the three basic families:
%%of twist $(N,1,1)$, double-braid $(N,M,L)$ and generic 3-pretzel knots $(N,M,L)$.
%%In particular, we obtain a beautiful formula for all symmetric??? representations,
%%which reveals the way in which the celebrated {\it factorization} of DE coefficients for
%%double braids is deformed.
%%This solves an old puzzle, but implications remain to be drawn.
%
%sec.\ref{summary} lists the properties of the defect, which we learned so far.
sec.\ref{conc} provides a short summary.
In this paper we consider defects only for symmetric representations,
generalizations to rectangular and non-rectangular cases require more details
\cite{KM3,KNTZ1, Morozov4,Morozov3}
and will be considered elsewhere.

\section{Differential expansion for symmetric representations $[r]$ 
\label{diffexp}}

(Reduced) HOMPLY-PT polynomials\footnote{We always omit the word "polynomials" for simplicity.} are analytically continued matrix elements or weighted traces of
various products of ${\cal R}$-matrices in particular representation $R$ and its conjugate $\bar R$ of $\mathfrak{sl}_N$. Accordingly they inherit the basic properties of the representation theory.
Translation language includes several rules:
\begin{itemize}
    \item HOMFLY-PT is a clever analytical continuation in $N$. Reduction to particular $N$, i.e to the quantum $\mathfrak{sl}_N$ invariants appears at $A=q^N$.
     \item For a given $N$ representation theory of $\mathfrak{sl}_N$ is applicable only for Young diagrams $R$ with no more than $N$ rows $l_R\leqslant N$. We do not discuss the "non-physical" region $l_R \geqslant N$, that appears to be rather interesting due to the presence of tug-the-hook symmetry and stability property \cite{MST1, MST2, KM1}.
    \item Transposition of the Young diagram $R\longrightarrow  R^{T}$ is equivalent to the substitution $q\longrightarrow q^{-1}$:
    \begin{equation}
     H^\mathcal{K}_R(A, q)=H^\mathcal{K}_{R^T}(A, q^{-1}), \hspace{15mm}
     \ytableausetup{boxsize=0.6em,aligntableaux = center}
     R=\ydiagram{4,2,1} \hspace{5mm}\longleftrightarrow \hspace{5mm}\ydiagram{3,2,1,1}=R^T.
    \end{equation}
    \item Conjugation $R\longrightarrow  \overline{R}$ is a symmetry of HOMFLY-PT. The symmetry depends on the rank $N$, we provide an example diagram for $N=5$:
\begin{equation}
\ytableausetup{boxsize=0.6em,aligntableaux = center}
 H^\mathcal{K}_R(A, q)=H^\mathcal{K}_{\overline{R}}(A, q), \hspace{15mm}
R=\begin{ytableau}
 \ & \ & \ & \ \\
 \ & \ & \ & *(gray) \\
 \ & *(gray) & *(gray) & *(gray) \\
  *(gray) & *(gray) & *(gray) & *(gray) \\
  *(gray) & *(gray) & *(gray) & *(gray) \\
\end{ytableau} \hspace{5mm}\longleftrightarrow \hspace{5mm}\begin{ytableau}
 *(gray) & *(gray) & *(gray) & *(gray) \\
 *(gray) & *(gray) & *(gray) & *(gray) \\
 *(gray) & *(gray) & *(gray) \\
 *(gray) \\
\end{ytableau}=\overline{R}
\end{equation}
\end{itemize}

Using these group-theoretical properties we {\it drastically} restrict the form of HOMFLY-PT for the symmetric representations $[r]$.
In what follows we use the standard abbreviations \footnote{We hope the abuse of notation for symmetric representations $[r]$ and quantum numbers $[k]_q$ does not cause a confusion.} $\{x\} := x-x^{-1}$ and $[k]_q:=\frac{\{q^k\}}{\{q\}}$.

For $N=1$ representation theory is trivial, this means that $H^{\mathcal{K}}_R(A=q,q)=1$,
i.e. 
\be
\label{r sym}
H^{\mathcal{K}}_R(A,q) - 1 \ \vdots \ \{A/q\}
\ee
but only for $l_R\leqslant 1$, i.e.
for symmetric representations $R=[r]$.
Also we have for $l_{R^{T}}\leqslant 1$, i.e. for anti-symmetric representations $R = [1^{r}]$:
\be
    H^{\mathcal{K}}_{R^{T}}(A,q) - 1 \ \vdots \ \{A/q\} \hspace{10mm} \Longrightarrow \hspace{10mm} H^{\mathcal{K}}_R(A,q)-1 \ \vdots \ \{Aq\}
\ee
Both  restrictions hold for the fundamental representation $R=[1]$
and therefore we have:
\be 
H^{\mathcal{K}}_{[1]}-1 \ \vdots \ \{Aq\}\{A/q\}
\ee
or alternatively
\be
H_{[1]}^{\cal K}(A,q) = 1 + \textbf{F}_{[1]}^{\cal K}(A,q)\cdot \{Aq\}\{A/q\}
\ee
with some function $\textbf{F}_{[1]}^{\cal K}(A,q)$.
This is the simplest example of differential expansion \cite{BM1,Morozov1, Morozov2,Morozov3,Morozov4, Morozov5, Morozov6, BJLMMMS, MMM1,IMMM, KM1,KM2,KM3}
(it is also known as cyclotomic expansion \cite{Habiro2007,Nawata:2015wya,KNTZ1,CLZ,Chen:2014fpa,Chen:2015rid,Kawagoe:2012bt,Kawagoe:2021onh,BG}).

Further we discuss first symmetric representation $[2]$ in detail while generic symmetric representation $[r]$ could be obtain with the same logic. For $N = 2$ the first anti-symmetric representation $R = [1,1]$ is trivial:
\be
H^{\mathcal{K}}_{[1,1]}(A,q) - 1  \ \vdots \ \{A / q^2\} \hspace{10mm} \Longrightarrow \hspace{10mm} H^{\mathcal{K}}_{[2]}(A,q)-1 \ \vdots \ \{Aq^2\}
\ee
Combining this result with \eqref{r sym} we get
\be
\label{2sym}
H^{\mathcal{K}}_{[2]}(A,q) - 1  \ \vdots \ \{A q^2\} \{A/q\} \hspace{10mm} \Longrightarrow \hspace{10mm} H^{\mathcal{K}}_{[2]}(A,q) = 1 + \{A q^2\} \{A/q\} \cdot T^{\mathcal{K}}_{[2]}(A,q)
\ee
with some auxiliary function $T_{[2]}^{\mathcal{K}}$. Next, we use relation from $N=3$ representation theory $[1,1] \approx [1]$:
\be
\label{3sym}
H^{\mathcal{K}}_{[1,1]}(A,q) - H^{\mathcal{K}}_{[1]}(A,q)  \ \vdots \ \{A /q^3\}  \hspace{10mm} \Longrightarrow \hspace{10mm} H^{\mathcal{K}}_{[2]}(A,q) - H^{\mathcal{K}}_{[1]}(A,q)  \ \vdots \ \{A q^3\} 
\ee
Resolving \eqref{2sym} and \eqref{3sym} with the help of identity $ [2]_q \{ A q^2 \} = \{ A q \} + \{ A q^3 \}$ we obtain:
\be
T_{[2]}^{\mathcal{K}}(A,q) - [2]_q \cdot \textbf{F}_{[1]}^{\mathcal{K}}(A,q)  \ \vdots \ \{A q^3\}  
\ee
or in alternative form
\be
H^{\mathcal{K}}_{[2]}(A,q) = 1 + [2]_q \cdot \textbf{F}_{[1]}^{\mathcal{K}}(A,q) \cdot \{A q^2\} \{A/q\} + \textbf{F}_{[2]}^{\mathcal{K}}(A,q) \cdot \{A q^3 \}\{A q^2\} \{A/q\}
\ee
for some knot-dependent function $\textbf{F}_{[2]}^{\mathcal{K}}(A,q)$.
For bigger symmetric representations we follow the same method as for $[2]$ and use the fact from the representation theory for anti-symmetric representations
$[1^r]\approx [1^{N-r}]$.
By induction one can {\it prove} the differential expansion (DE) formula for generic symmetric representation $[r]$ \cite{IMMM} 
\be
\boxed{
H^{\cal K}_{[r]}(A,q) = 1
+ \sum_{s=1}^r \, \frac{[r]_q!}{[s]_q! [r-s]_q!} \cdot {\bf F}_{[s]}^{\cal K}(A,q)\cdot
\{A/q\}\prod_{i=0}^{s-1} \{Aq^{r+i}\}
\label{basicDE}
}
\ee

For non-symmetric $R$ the story gets more involved, see \cite{Morozov1,Morozov2,Morozov3,Morozov4,Morozov5,KM2,KM3,KNTZ1}. DE formula \eqref{basicDE} shows that the topological information is carried only by DE coefficients $\textbf{F}_{[s]}^{\mathcal{K}}(A,q)$, while the other constituents of the HOMFLY-PT do not help in distinguishing knots. 

Interestingly, for every symmetric representations $[r]$ there are exactly $r$ DE coefficients, most of them $\textbf{F}_{[s]}^{\mathcal{K}}(A,q) \ s = 1,\ldots,r-1$ are came from smaller representations and only one $\textbf{F}_{[r]}^{\mathcal{K}}(A,q)$ is new.

\section{Defect of the differential expansion
\label{defects}}

\subsection{The basic idea and the ladder structure
\label{ladders}}

For generic knots ${\cal K}$ expansion (\ref{basicDE}) is the best one can achieve from the naive group-theoretical reasoning.
However, sometime, e.g. for figure-eight knot $4_1$ or a trefoil $3_1$,
it can be significantly enhanced:
\be
H^{{\cal K}_0}_{[r]}(A,q) = 1
+ \sum_{s=1}^r   \frac{[r]_q!}{[s]_q! [r-s]_q!} \cdot
{\cal F}_{[s]}^{{\cal K}_0}(A,q)\cdot
\prod_{i=0}^{s-1} \{Aq^{r+i}\}\{Aq^{i-1}\}
\label{def0DE}
\ee
in other words
\be
{\bf F}^{{\cal K}_0}_{[s]} = {\cal F}^{{\cal K}_0}_{[s]}\cdot \prod_{i=1}^{s-1} \{Aq^{i-1}\}
\ee
the DE coefficients are further factorized for particular knots.
The knots ${\cal K}_0$ with this property
are said to have {\it defect zero} \cite{KM1}, $\delta^{\cal K}=0$.
For a generic knot the factorization of ${\bf F}^{\cal K}$ is only partial,
\be
{\bf F}_{[s]}^{{\cal K}_\delta}(A,q)
= {\cal F}_{[s]}^{{\cal K}_\delta}(A,q)\cdot\prod_{i=1}^{\nu_s^\delta}\{Aq^{i-1}\}
\label{extrafactor}
\ee
and {\it defect} $\delta$ can take any non-negative integer value \cite{KM1},
with 
\be
\nu_s^{\delta} = {\rm entier}\left(\frac{s-1}{\delta+1}\right)
\label{ladderdefect}
\ee
These formulas illustrate a conjecture, that {\bf for a knot ${\cal K}$ one can identify one non-negative number $\delta^{{\cal K}}$ measuring factorization of the DE coefficients}. While this conjecture appears to be true in numerous examples, we observe certain anomalies for knots with unit Alexander polynomial and discuss it in sec.\ref{defAlex}.

The number of additional brackets \eqref{ladderdefect} can be represented as a peculiar ladder diagrams:
\be
\label{def 0 diag}
\begin{picture}(300,70)(-40,-30)
\put(0,-20){
\put(-90,0){\mbox{defect $\delta^{\cal K}=0$:}}
\put(-100,0){
\put(190,-20){\line(1,0){110}}
\put(190,-5){\line(1,0){110}}
\put(215,10){\line(1,0){85}}
\put(240,25){\line(1,0){60}}
\put(265,40){\line(1,0){35}}
\put(290,55){\line(1,0){10}}
%\put(240,40){\line(1,0){15}}
%\put(250,50){\line(1,0){5}}
%
\put(190,-20){\line(0,1){15}}
\put(215,-20){\line(0,1){30}}
\put(240,-20){\line(0,1){45}}
\put(265,-20){\line(0,1){60}}
\put(290,-20){\line(0,1){75}}
%\put(240,-20){\line(0,1){60}}
%\put(250,-20){\line(0,1){70}}
%
\put(194,-16){\mbox{\footnotesize $\{A\}$}}
\put(219,-16){\mbox{\footnotesize $\{A\}$}}
\put(244,-16){\mbox{\footnotesize $\{A\}$}}
\put(269,-16){\mbox{\footnotesize $\{A\}$}}
%\put(294,-16){\mbox{\footnotesize $\{A\}$}}
%\put(251,-16){\mbox{\tiny \{A\}}}
\put(218,-1){\mbox{\footnotesize $\{Aq\}$}}
\put(243,-1){\mbox{\footnotesize $\{Aq\}$}}
\put(268,-1){\mbox{\footnotesize $\{Aq\}$}}
%\put(293,-1 ){\mbox{\footnotesize $\{Aq\}$}}
\put(241,14){\mbox{\footnotesize $\{Aq^2\}$}}
\put(266,14){\mbox{\footnotesize $\{Aq^2\}$}}
%\put(291,14){\mbox{\footnotesize $\{Aq^2\}$}}
\put(266,29){\mbox{\footnotesize $\{Aq^3\}$}}
%\put(291,29){\mbox{\footnotesize $\{Aq^3\}$}}
%\put(291,44){\mbox{\footnotesize $\{Aq^4\}$}}
%
\put(170,-35){\mbox{$s$}}
\put(200,-35){\mbox{{\footnotesize $2$}}}
\put(225,-35){\mbox{{\footnotesize $3$}}}
\put(250,-35){\mbox{{\footnotesize $4$}}}
\put(275,-35){\mbox{{\footnotesize $5$}}}
\put(300,-35){\mbox{{\footnotesize $6$}}}
%\put(243,-35){\mbox{{\footnotesize $7$}}}
%\put(253,-35){\mbox{{\footnotesize $8$}}}
%\put(263,-35){\mbox{{\footnotesize $9$}}}
%\put(270,-35){\mbox{{\footnotesize $10$}}}
%\put(280,-35){\mbox{{\footnotesize $11$}}}
%\put(290,-35){\mbox{{\footnotesize $12$}}}
%\put(300,-35){\mbox{{\footnotesize $13$}}}
%\put(310,-35){\mbox{{\footnotesize $14$}}}
%
%\put(-50,0){
\put(320,-15){\mbox{{\footnotesize $1$}}}
\put(320,0){\mbox{{\footnotesize $2$}}}
\put(320,15){\mbox{{\footnotesize $3$}}}
\put(320,30){\mbox{{\footnotesize $4$}}}
\put(320,45){\mbox{{\footnotesize $5$}}}
%\put(320,32){\mbox{{\footnotesize $6$}}}
%\put(320,42){\mbox{{\footnotesize $7$}}}
%
%}
%
\put(170,31){\mbox{$\nu_s^{\cal K} = s-1$}}
}}
\end{picture}
\ee

\be
\begin{picture}(300,70)(-40,-20)
\put(0,-20){
\put(-90,0){\mbox{defect $\delta^{\cal K}=1$:}}
\put(-100,0){
\put(200,-20){\line(1,0){115}}
\put(200,-10){\line(1,0){115}}
\put(220,0){\line(1,0){95}}
\put(240,10){\line(1,0){75}}
\put(260,20){\line(1,0){55}}
\put(280,30){\line(1,0){35}}
\put(300,40){\line(1,0){15}}
\put(200,-20){\line(0,1){10}}
\put(210,-20){\line(0,1){10}}
\put(220,-20){\line(0,1){20}}
\put(230,-20){\line(0,1){20}}
\put(240,-20){\line(0,1){30}}
\put(250,-20){\line(0,1){30}}
\put(260,-20){\line(0,1){40}}
\put(270,-20){\line(0,1){40}}
\put(280,-20){\line(0,1){50}}
\put(290,-20){\line(0,1){50}}
\put(300,-20){\line(0,1){60}}
\put(310,-20){\line(0,1){60}}
%\put(320,-20){\line(0,1){70}}
%
\put(180,-35){\mbox{$s$}}
%\put(193,-35){\mbox{{\footnotesize $2$}}}
\put(203,-35){\mbox{{\footnotesize $3$}}}
\put(213,-35){\mbox{{\footnotesize $4$}}}
\put(223,-35){\mbox{{\footnotesize $5$}}}
\put(233,-35){\mbox{{\footnotesize $6$}}}
\put(243,-35){\mbox{{\footnotesize $7$}}}
\put(253,-35){\mbox{{\footnotesize $8$}}}
\put(263,-35){\mbox{{\footnotesize $9$}}}
\put(270,-35){\mbox{{\footnotesize $10$}}}
\put(280,-35){\mbox{{\footnotesize $11$}}}
\put(290,-35){\mbox{{\footnotesize $12$}}}
\put(300,-35){\mbox{{\footnotesize $13$}}}
\put(310,-35){\mbox{{\footnotesize $14$}}}
\put(320,-18){\mbox{{\footnotesize $1$}}}
\put(320,-8){\mbox{{\footnotesize $2$}}}
\put(320,2){\mbox{{\footnotesize $3$}}}
\put(320,12){\mbox{{\footnotesize $4$}}}
\put(320,22){\mbox{{\footnotesize $5$}}}
\put(320,32){\mbox{{\footnotesize $6$}}}
%\put(320,42){\mbox{{\footnotesize $7$}}}
%
\put(170,31){\mbox{$\nu_s^{\cal K} = \text{entier}\left( \frac{s-1}{2}\right)$}}
}}
\end{picture}
\ee

\be
\begin{picture}(300,80)(-40,0)
\put(0,-20){
\put(-90,20){\mbox{defect $\delta^{\cal K}=2$:}}
\put(-100,30){
\put(210,-20){\line(1,0){115}}
\put(210,-10){\line(1,0){115}}
\put(240,0){\line(1,0){85}}
\put(270,10){\line(1,0){55}}
\put(300,20){\line(1,0){25}}
%\put(280,30){\line(1,0){35}}
%\put(300,40){\line(1,0){15}}
%
%\put(200,-20){\line(0,1){10}}
\put(210,-20){\line(0,1){10}}
\put(220,-20){\line(0,1){10}}
\put(230,-20){\line(0,1){10}}
\put(240,-20){\line(0,1){20}}
\put(250,-20){\line(0,1){20}}
\put(260,-20){\line(0,1){20}}
\put(270,-20){\line(0,1){30}}
\put(280,-20){\line(0,1){30}}
\put(290,-20){\line(0,1){30}}
\put(300,-20){\line(0,1){40}}
\put(310,-20){\line(0,1){40}}
\put(320,-20){\line(0,1){40}}
\put(190,-35){\mbox{$s$}}
%\put(193,-35){\mbox{{\footnotesize $2$}}}
%\put(203,-35){\mbox{{\footnotesize $3$}}}
\put(213,-35){\mbox{{\footnotesize $4$}}}
\put(223,-35){\mbox{{\footnotesize $5$}}}
\put(233,-35){\mbox{{\footnotesize $6$}}}
\put(243,-35){\mbox{{\footnotesize $7$}}}
\put(253,-35){\mbox{{\footnotesize $8$}}}
\put(263,-35){\mbox{{\footnotesize $9$}}}
\put(270,-35){\mbox{{\footnotesize $10$}}}
\put(280,-35){\mbox{{\footnotesize $11$}}}
\put(290,-35){\mbox{{\footnotesize $12$}}}
\put(300,-35){\mbox{{\footnotesize $13$}}}
\put(310,-35){\mbox{{\footnotesize $14$}}}
\put(330,-18){\mbox{{\footnotesize $1$}}}
\put(330,-8){\mbox{{\footnotesize $2$}}}
\put(330,2){\mbox{{\footnotesize $3$}}}
\put(330,12){\mbox{{\footnotesize $4$}}}
%\put(320,22){\mbox{{\footnotesize $5$}}}
%\put(320,32){\mbox{{\footnotesize $6$}}}
%\put(320,42){\mbox{{\footnotesize $7$}}}
%
\put(195,25){\mbox{$\nu_s^{\cal K}=\text{entier}\left( \frac{s-1}{3}\right)$}}
}}
\end{picture}
\ee

\be
\begin{picture}(300,70)(-40,20)
\put(0,-20){
\put(-90,40){\mbox{defect $\delta^{\cal K}=3$:}}
\put(-100,50){
\put(220,-20){\line(1,0){105}}
\put(220,-10){\line(1,0){105}}
\put(260,0){\line(1,0){65}}
\put(300,10){\line(1,0){25}}
%\put(310,20){\line(1,0){55}}
%\put(280,30){\line(1,0){35}}
%\put(300,40){\line(1,0){15}}
%
%\put(200,-20){\line(0,1){10}}
%\put(210,-20){\line(0,1){10}}
\put(220,-20){\line(0,1){10}}
\put(230,-20){\line(0,1){10}}
\put(240,-20){\line(0,1){10}}
\put(250,-20){\line(0,1){10}}
\put(260,-20){\line(0,1){20}}
\put(270,-20){\line(0,1){20}}
\put(280,-20){\line(0,1){20}}
\put(290,-20){\line(0,1){20}}
\put(300,-20){\line(0,1){30}}
\put(310,-20){\line(0,1){30}}
\put(320,-20){\line(0,1){30}}
\put(200,-35){\mbox{$s$}}
%\put(193,-35){\mbox{{\footnotesize $2$}}}
%\put(203,-35){\mbox{{\footnotesize $3$}}}
%\put(213,-35){\mbox{{\footnotesize $4$}}}
\put(223,-35){\mbox{{\footnotesize $5$}}}
\put(233,-35){\mbox{{\footnotesize $6$}}}
\put(243,-35){\mbox{{\footnotesize $7$}}}
\put(253,-35){\mbox{{\footnotesize $8$}}}
\put(263,-35){\mbox{{\footnotesize $9$}}}
\put(270,-35){\mbox{{\footnotesize $10$}}}
\put(280,-35){\mbox{{\footnotesize $11$}}}
\put(290,-35){\mbox{{\footnotesize $12$}}}
\put(300,-35){\mbox{{\footnotesize $13$}}}
\put(310,-35){\mbox{{\footnotesize $14$}}}
\put(330,-18){\mbox{{\footnotesize $1$}}}
\put(330,-8){\mbox{{\footnotesize $2$}}}
\put(330,2){\mbox{{\footnotesize $3$}}}
%\put(320,12){\mbox{{\footnotesize $4$}}}
%\put(320,22){\mbox{{\footnotesize $5$}}}
%\put(320,32){\mbox{{\footnotesize $6$}}}
%\put(320,42){\mbox{{\footnotesize $7$}}}
%
\put(200,17){\mbox{$\nu_s^{\cal K}=\text{entier}\left( \frac{s-1}{4}\right)$}}
}}
\end{picture}
\ee

\be
\ldots
\nn
\ee
Each box in these diagram correspond to the particular brackets as in \eqref{def 0 diag}. 
We provide values of defect for several simple knot families. For antiparallel\footnote{The overlined $\overline{N}$ items means antiparallel braid in the pretzel notation.} pretzel knots $(\overline{N_1},\ldots, \overline{N_k})$ for odd $N_i$ and odd $k$, where $N_i$ are the numbers of crossings in each 2-strand braid :
\be
\label{pretzdef}
\delta^{(\overline{N_1},\ldots, \overline{N_k})} = \frac{k-3}{2}
\ee
This pretzel family contains all twist $(N_1,1,1)$ and double braid $(N_1, N_2, 1)$ knots for which defect vanishes. Another family of torus knots $T[M,N]$ for $M,N >0$ have the following defects:
\be
\delta^{T[M,N]} = \frac{M N - M - N - 1}{2}
\ee
More complicated examples are discussed in sec.\ref{compexa}.

Every particular knot ${\cal K}$ has a certain defect,
and this means that its DE is obligatory enhanced
as compared to (\ref{basicDE}).
However, there are knots with arbitrarily large defects
and additional factorization of ${\bf F}^{\cal K} $ can be made as weak as one wishes by appropriate selection of ${\cal K}$.

For the unknot defect is not well defined:
all the coefficients ${\bf F}^{\rm unknot}  = 0$,
i.e. one can prescribe any degree of factorization to them.
Since (\ref{ladderdefect}) has  an apparent singularity at $\delta=-1$,
it looks natural to put
\be
\delta^{\rm unknot}\ \stackrel{?}{=}\  -1
\label{unknotdefect}
\ee
Indeed, this is often the implication of evolution formulas
for families, which involve unknots at particular
values of evolution parameter.

%Remarkably, there is also a stronger option $\delta^{\cal K}=-1$, when
%factorization is further {\it enhanced} to fill the remaining gap:
%\be
%H^{{\cal K}_{-1}}_{[r]}(A,q) = 1
%+ \sum_{j=1}^r {\cal F}_{[j]}^{{\cal K}_{-1}}(A,q)\cdot
%\overbrace{\prod_{i=0}^{j-1} \{Aq^{r+i}\}\prod_{i=0}^j \{Aq^{i-1}\}}^{{\cal Z}_{[r]}^{[i]}}
%\label{def0DE}
%\ee
%i.e.  $F^{{\cal K}_{-1}}_{[j]} = {\cal F}^{{\cal K}_{-1}}_{[j]}\cdot \{Aq^{j-1}\}$,
%so that the highest item with $j=r$ in the sum has {\it no gap}:
%the highest $Z$-factor becomes ${\cal Z}_{[r]}^{[r]} = \prod_{i=0}^{2r} \{Aq^{i-1}\}$.

\subsection{Evolution of defect}

Analysis of various examples,
which will be partly described in the text below,
leads us to the following hypothesis that we call {\it defect evolution hypothesis}:

\bigskip
\begin{itemize}
    \item  {\bf Defect does not change when any vertex is substituted by an odd antiparallel braid
of any length}.

    \item {\bf When the length of any parallel braid is increased by two, defect changes by one}.
\end{itemize}
\bigskip
In other words, we can pick up any vertex in the knot graph and substitute it by
a triple -- in two possible directions.
The conjecture is then equivalent to the following picture:

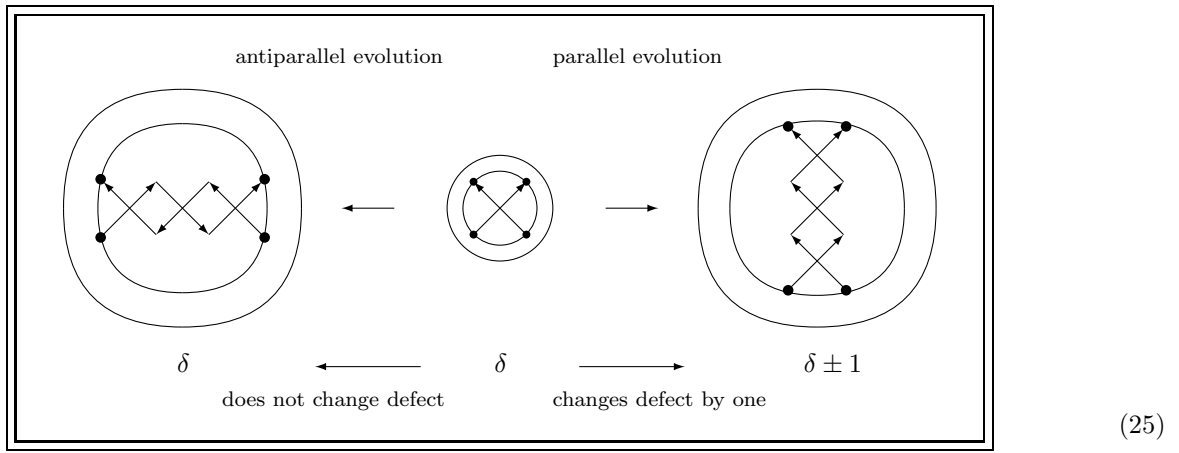
\be
\boxed{
\boxed{
\begin{picture}(360,155)(-180,-85)

\put(-10,-10){\vector(1,1){20}}
\put(10,-10){\vector(-1,1){20}}
\put(0,0){\circle{28}}
\put(0,0){\circle{40}}

\put(-10,-10){\circle*{3}}
\put(-10,10){\circle*{3}}
\put(10,-10){\circle*{3}}
\put(10,10){\circle*{3}}

\put(40,0){\vector(1,0){20}}
\put(-40,0){\vector(-1,0){20}}
\put(-2,-62){\mbox{$\delta $}}
\put(30,-60){\vector(1,0){40}}
\put(-30,-60){\vector(-1,0){40}}
\put(20,55){\mbox{\text{\footnotesize parallel evolution}}}
\put(-100,55){\mbox{\text{\footnotesize antiparallel evolution}}}
\put(20,-75){\mbox{\text{\footnotesize changes defect by one}}}
\put(-105,-75){\mbox{\text{\footnotesize does not change defect}}}

\put(120,0){
\put(-10,-30){\vector(1,1){20}}
\put(10,-30){\vector(-1,1){20}}
\put(-10,-10){\vector(1,1){20}}
\put(10,-10){\vector(-1,1){20}}
\put(-10,10){\vector(1,1){20}}
\put(10,10){\vector(-1,1){20}}

\qbezier(0,33)(33,33)(33,0)
\qbezier(0,33)(-33,33)(-33,0)
\qbezier(0,-33)(33,-33)(33,0)
\qbezier(0,-33)(-33,-33)(-33,0)

\qbezier(0,45)(45,45)(45,0)
\qbezier(0,45)(-45,45)(-45,0)
\qbezier(0,-45)(45,-45)(45,0)
\qbezier(0,-45)(-45,-45)(-45,0)

\put(-11,-31){\circle*{4}}
\put(-11,31){\circle*{4}}
\put(11,-31){\circle*{4}}
\put(11,31){\circle*{4}}

\put(-5,-62){\mbox{$\delta \pm 1 $}}
}

\put(-120,0){
\put(-30,-10){\vector(1,1){20}}
\put(-10,10){\vector(1,-1){20}}
\put(10,-10){\vector(1,1){20}}
\put(30,-10){\vector(-1,1){20}}
\put(10,10){\vector(-1,-1){20}}
\put(-10,-10){\vector(-1,1){20}}

\qbezier(0,32)(32,32)(32,0)
\qbezier(0,32)(-32,32)(-32,0)
\qbezier(0,-32)(32,-32)(32,0)
\qbezier(0,-32)(-32,-32)(-32,0)

\qbezier(0,45)(45,45)(45,0)
\qbezier(0,45)(-45,45)(-45,0)
\qbezier(0,-45)(45,-45)(45,0)
\qbezier(0,-45)(-45,-45)(-45,0)

\put(-31,-11){\circle*{4}}
\put(-31,11){\circle*{4}}
\put(31,-11){\circle*{4}}
\put(31,11){\circle*{4}}

\put(-2,-62){\mbox{$\delta $}}
}
\end{picture}
}}
\label{conjecture}
\ee

\bigskip

Two additional comments make the statement more accurate:

\bigskip

-- Note that the length is a modulus of the evolution parameter, thus there is a non-analyticity when parameter changes sign. We explain this fact on a particular examples in sec.\ref{2torus} and sec.\ref{compexa}.

-- In the case of anti-parallel evolution this non-analyticity is reflected in a possible drop of the defect by unity for {\it one} particular length of the anti-parallel braid, sec.\ref{compexa}.

\bigskip

Changing of one vertex for two rather than three
converts a knot into a link and is not considered in this paper
(we remind that we deal with {\it reduced} HOMFLY-PT, which are defined differently
for knots and links).

\subsection{Defect and Alexander polynomial
\label{defAlex}}

As conjectured in the original paper \cite{KM1},
defect $\delta_{\cal K}$ is related to the degree of the fundamental Alexander polynomial
in $q^{\pm 1}$:
\be
{\rm Al}^{\cal K}_{[1]} = 1 + \{q\}^2 \cdot {\rm Pol}^{{\cal K}_\delta } (q^2,q^{-2})
\label{Aldef}
\ee
where  ${\rm Pol}^{{\cal K}_\delta}$ is a Laurent polynomial in $q^2$ of {\it degree $\delta^{\cal K}$}, symmetric under the
change $q^2\leftrightarrow q^{-2}$. This observation could be used as an "alternative definition" of defect.
In particular, for $\delta^{\cal K}=0$
the coefficient of $\{q\}^2$ in the fundamental Alexander reduces to a constant.

The quantity ${\rm Pol}^{{\cal K}_\delta}$ depends on the knot, and sometimes it can vanish,
so that ${\rm Al}_{[1]} = 1$.
In \cite{KM1} it was suggested to treat this situation as
defect $\delta=-1$, but this turns out to be a {\bf wrong idea}.
We suggest to substitute it by a more viable alternative --
that defect is a property of an {\it evolution family},
and some coefficients ${\bf F}_{[s]}$ can "accidently" factorize further
at some particular values of the evolution parameters.
%???
Notably, in all the examples additional factorization does not break
the ladder structure, just extends the list of allowed ladders
over the one in sec.\ref{ladders}.
As to $\delta=-1$, we now reserve this value to the unknot only,
in accordance with (\ref{unknotdefect}).

\bigskip

Since (\ref{Aldef}) involves only the fundamental representation,
in this case we can apply HOMFLY-PT skein relation
\be
%???A\ X -A^{-1}\ X^{-1} = (q-q^{-1})\ || ????
\begin{picture}(300,25)(-70,-30)

\put(0,-30){
\put(-25,10){\mbox{$A $}}
\put(-10,0){\vector(1,1){24}}
\put(14,0){\line(-1,1){10}}
\put(0,14){\vector(-1,1){10}}

\put(60,0){
\put(-38,10){\mbox{$= \ \ A^{-1} $}}
\put(24,0){\vector(-1,1){24}}
\put(0,0){\line(1,1){10}}
\put(14,14){\vector(1,1){10}}
}

\put(135,0){
\put(-39,10){\mbox{$+ \ \ \{q\}\ \cdot $}}
\put(0,0){\vector(0,1){24}}
\put(10,0){\vector(0,1){24}}
}
}

\end{picture}
\ee
to (\ref{conjecture}) at the particular point $A = 1$ reducing HOMFLY-PT to the Alexander polynomial. 
Then we can substitute the evolution hypothesis for a relation
between Alexander polynomials for a knot and a pair of associated links:

\be
\begin{picture}(360,155)(-180,-85)

\put(-10,-10){\vector(1,1){20}}
\put(10,-10){\vector(-1,1){20}}
\put(0,0){\circle{28}}
\put(0,0){\circle{40}}

\put(-10,-10){\circle*{3}}
\put(-10,10){\circle*{3}}
\put(10,-10){\circle*{3}}
\put(10,10){\circle*{3}}

\put(40,0){\mbox{+}}
\put(60,0){\mbox{$\{q\}\cdot$}}
\put(-50,0){\mbox{ =}}
\put(-2,-62){\mbox{$\delta $}}
%\put(30,-60){\vector(1,0){40}}
\put(-30,-60){\vector(-1,0){40}}
\put(-70,55){\mbox{\text{\footnotesize antiparallel evolution
does not change defect}}}
%\put(-60,-75){\mbox{\text{\footnotesize does not change defect}}}

\put(100,0){
\qbezier(-10,-10)(0,0)(-10,10)
\qbezier(10,-10)(0,0)(10,10)

\put(0,0){\circle{28}}
\put(0,0){\circle{40}}

\put(-10,-10){\circle*{3}}
\put(-10,10){\circle*{3}}
\put(10,-10){\circle*{3}}
\put(10,10){\circle*{3}}
}

\put(-120,0){
\put(-30,-10){\vector(1,1){20}}
\put(-10,10){\vector(1,-1){20}}
\put(10,-10){\vector(1,1){20}}
\put(30,-10){\vector(-1,1){20}}
\put(10,10){\vector(-1,-1){20}}
\put(-10,-10){\vector(-1,1){20}}

\qbezier(0,32)(32,32)(32,0)
\qbezier(0,32)(-32,32)(-32,0)
\qbezier(0,-32)(32,-32)(32,0)
\qbezier(0,-32)(-32,-32)(-32,0)

\qbezier(0,45)(45,45)(45,0)
\qbezier(0,45)(-45,45)(-45,0)
\qbezier(0,-45)(45,-45)(45,0)
\qbezier(0,-45)(-45,-45)(-45,0)

\put(-31,-11){\circle*{4}}
\put(-31,11){\circle*{4}}
\put(31,-11){\circle*{4}}
\put(31,11){\circle*{4}}

\put(-2,-62){\mbox{$\delta $}}
}
\end{picture}
\ee

\be
\begin{picture}(360,155)(-180,-85)

\put(-10,-10){\vector(1,1){20}}
\put(10,-10){\vector(-1,1){20}}
\put(0,0){\circle{28}}
\put(0,0){\circle{40}}

\put(-10,-10){\circle*{3}}
\put(-10,10){\circle*{3}}
\put(10,-10){\circle*{3}}
\put(10,10){\circle*{3}}

\put(40,0){\mbox{+}}
\put(-50,0){\mbox{=}}
\put(-2,-62){\mbox{$\delta $}}
\put(60,0){\mbox{$\{q\}\ \cdot$}}
%\put(30,-60){\vector(1,0){40}}
\put(-30,-60){\vector(-1,0){40}}
\put(-60,55){\mbox{\text{\footnotesize parallel evolution changes defect by one}}}
%\put(-100,55){\mbox{\text{\footnotesize antiparallel evolution}}}
%\put(20,-75){\mbox{\text{\footnotesize changes defect by one}}}
%\put(-105,-75){\mbox{\text{\footnotesize does not change defect}}}

\put(-120,0){
\put(-10,-30){\vector(1,1){20}}
\put(10,-30){\vector(-1,1){20}}
\put(-10,-10){\vector(1,1){20}}
\put(10,-10){\vector(-1,1){20}}
\put(-10,10){\vector(1,1){20}}
\put(10,10){\vector(-1,1){20}}

\qbezier(0,33)(33,33)(33,0)
\qbezier(0,33)(-33,33)(-33,0)
\qbezier(0,-33)(33,-33)(33,0)
\qbezier(0,-33)(-33,-33)(-33,0)

\qbezier(0,45)(45,45)(45,0)
\qbezier(0,45)(-45,45)(-45,0)
\qbezier(0,-45)(45,-45)(45,0)
\qbezier(0,-45)(-45,-45)(-45,0)

\put(-11,-31){\circle*{4}}
\put(-11,31){\circle*{4}}
\put(11,-31){\circle*{4}}
\put(11,31){\circle*{4}}

\put(-5,-62){\mbox{$\delta+1 $}}
}

\put(120,0){
\put(-10,-20){\vector(1,1){20}}
\put(10,-20){\vector(-1,1){20}}
\put(-10,-0){\vector(1,1){20}}
\put(10,-0){\vector(-1,1){20}}
%\put(-10,10){\vector(1,1){20}}
%\put(10,10){\vector(-1,1){20}}

\qbezier(0,23)(23,23)(23,0)
\qbezier(0,23)(-23,23)(-23,0)
\qbezier(0,-23)(23,-23)(23,0)
\qbezier(0,-23)(-23,-23)(-23,0)

\qbezier(0,35)(35,35)(35,0)
\qbezier(0,35)(-35,35)(-35,0)
\qbezier(0,-35)(35,-35)(35,0)
\qbezier(0,-35)(-35,-35)(-35,0)

\put(-11,-21){\circle*{4}}
\put(-11,21){\circle*{4}}
\put(11,-21){\circle*{4}}
\put(11,21){\circle*{4}}

}
\end{picture}
\ee
%??????
The Alexander skein relations appear to be useful in testing the defect evolution hypothesis. For the anti-parallel evolution, the hypothesis is true if the second term has a smaller or equal degree than the first term on the r.h.s. Similarly, for the parallel evolution, the second term should have greater degree than the first term on the r.h.s.

\subsection{Stability of unphysical $H_{[1^r]}(A=q^m)$ for a given defect
\label{stabil}}

%???

As a corollary of differential expansion,
HOMFLY-PT at $A=q^{-m}$
possesses a remarkable stability property \cite{KM1}:
its coefficients do not change with the increase of the
representation $[r]$ if $r$ is big enough as compared to $m\cdot \delta^{\cal K}$.
This is deep in unphysical domain, where the rank of the group exceeds the
number of lines in the Young diagram, still reduced HOMFLY-PT remain well defined.
We do not deal with this stabilization in the present paper, thus do not
formulate it in details, just mention it in the list of defect properties
in the conclusion.

%??????

\section{Basic examples
\label{basexa}}

%\section{Examples of antiparallel evolution}

\subsection{Triple anti-parallel pretzels have defect 0
\label{3pretz}}

%???

In this section we analyze the family of
triple antiparallel pretzels $(\overline{N},\overline{M},\overline{L})$.
%which will be discussed in more detail in sec.\ref{3pretzels}???.
To get a knot (rather then link) all the three parameters $N,M,L$ should be odd.
The family contains the previously studied twist and double braid
knots as particular cases $(\overline{N},1,1)$ and $(\overline{N},\overline{M},1)$.
The whole family can be considered as a triple antiparallel evolution of the trefoil:
each of the three ${\cal R}$-matrices are substituted by an antiparallel 2-strand braid.
%???picture%

According to \cite{MMMRS} colored HOMFLY-PT for antiparallel pretzels in symmetric representations $R=[r]$ could be calculated by the following arborescent formula:
\be
H_R^{(\overline{N},\overline{M},\overline{L})} =\  d_R^2\!\!  \sum_{X \in R\otimes R}\!\!
%d_X^{-1/2}
\frac{\left(\bar S\bar T^{N}S\right)_{\emptyset X}
\left(\bar S\bar T^{M}S\right)_{\emptyset X}\left(\bar S\bar T^{L}S\right)_{\emptyset X}}
{\sqrt{d_X}}
%= \nn \\
=  \ d_R^2\!\! \sum_{X\in R\otimes \bar R}\!\!
\frac{{\cal A}^{\rm{ea}}_{1X}(\overline{N}){\cal A}^{\rm{ea}}_{1X}(\overline{M})
{\cal A}^{\rm{ea}}_{1X}(\overline{L})}{\sqrt{d_X}}
\label{Hnml}
\ee
%
%\be
%H_R^{(2n-1,2m-1,2l-1)} = d_R^2 \sum_{X \in R\otimes R}
%%d_X^{-1/2}
%\frac{\left(\bar S\bar T^{2n-1}S\right)_{\emptyset X}
%\left(\bar S\bar T^{2m-1}S\right)_{\emptyset X}\left(\bar S\bar T^{2l-1}S\right)_{\emptyset X}}
%{\sqrt{d_X}}
%= \nn \\
%=  d_R^2\sum_{X\in R\otimes \bar R}
%\frac{{\cal A}^{\rm{ea}}_{1X}(\overline{2n-1}){\cal A}^{\rm{ea}}_{1X}(\overline{2m-1})
%{\cal A}^{\rm{ea}}_{1X}(\overline{2l-1})}{\sqrt{d_X}}
%\label{Hnml}
%\ee
where $d_X$ are quantum dimensions of representations $X$.
In the last expression we used the generic notation for pretzel fingers from sec.\ref{prefing} below.

All these knots have defect zero, thus the differential expansion is
\begin{align}
\begin{aligned}
H_{[1]}^{(\overline{N},\overline{M},\overline{L})} &= 1 + {\cal F}_{[1]}^{(\overline{N},\overline{M},\overline{L})}(A)\cdot \{Aq\}\{A/q\}  \\
H_{[2]}^{(\overline{N},\overline{M},\overline{L})} &= 1 + [2] \cdot {\cal F}_{[1]}^{(\overline{N},\overline{M},\overline{L})}(A)\cdot \{Aq^2\}\{A/q\}
+ {\cal F}_{[2]}^{(\overline{N},\overline{M},\overline{L})}(A,q)\cdot \{Aq^3\}\{Aq^2\}\{A\}\{A/q\}
 \\
\ldots  \\
H_{[r]}^{(\overline{N},\overline{M},\overline{L})} &= 1+ \sum_{s=1}^r \frac{[r]!}{[s]![r-s]!} \cdot {\cal F}_{[s]}^{(\overline{N},\overline{M},\overline{L})} \cdot
\prod_{j=0}^{s-1} \{Aq^{r+j}\}\{Aq^{j-1}\}
\end{aligned}
\end{align}
%where we use the standard notation $\{x\} = x-x^{-1}$ and $[k]:=\frac{\{q^k\}}{\{q\}}$
This is the simplest illustration -- and, actually, the origin -- of our hypothesis
that {\bf defect does not change with the antiparallel evolution}.

It is instructive to comment on the relation with "alternative definition" of defect in the
sec.\ref{defAlex} -- to better illustrate and understand its limitation.
We will do this in sec.\ref{3pretAlex}.

%\section{Examples of parallel evolution}

\subsection{Antiparallel pretzels as descendants of 2-strand torus knots
\label{2torus}}

HOMFLY-PT for the 2-strand torus knots are just traces of
${\cal R}$-matrices for the parallel 2-strand braid,
\be
H^{T[2n-1,2]}_R = \Tr\!_R S^{2n-1}
\ee
and can be calculated by a variety of methods \cite{DMMSS}.
It is easy to check that they have defects
\be
\delta^{T[2n-1,2]}
%= \left|n-\frac{1}{2}\right|-\frac{3}{2}
= \left\{
\begin{array}{ccc}
n-2 & {\rm for} & n\geq 2 \\  -n-1 &{\rm for} & n\leq -1
\end{array}
\right.
= \left|n-\frac{1}{2}\right|-\frac{3}{2}
= \frac{|2n-1|-3}{2}
\label{2torusdefect}
\ee
($n=0,1$ correspond to the unknot, when defect is not well defined),
in particular the trefoil, which we get at $n=- 1,2$ has defect zero.
In the previous subsection we considered trefoil as a starting point
for triple antiparallel evolution,
now it is an origin of a single parallel one.

The formula (\ref{2torusdefect}) is precise formulation of
the parallel part of our conjecture:
{\bf parallel evolution increases defect by one per pair of added vertices}.
It also reflects the main subtlety of this formulation:
the break of analyticity through occurrence of absolute value
and the presence of a "blind zone" at $n=0,1$.
The structure of these blind zones will get more sophisticated in
more general examples.

A somewhat similar break of analyticity takes place \cite{Anokhina:2018ysw,AMP,Morozov:2018ges}
in evolution of Khovanov and super-polynomials --
it would be interesting to establish a more clear relation between the two.

This example \eqref{2torusdefect} also nicely explains the defects of odd antiparallel pretzels \eqref{pretzdef}: they are obtained by the antiparallel evolution from 2-strand torus knots.

\section{Comments on the fundamental Alexander
\label{3pretAlex}}

Calculations of the defect could be drastically simplified,
if one used the "alternative definition" of defect in the sec.\ref{defAlex} --
only Alexander polynomial in the very first fundamental representation
would be needed.
This is indeed very helpful in majority of cases,
but unfortunately this method is not truly reliable.
Sometime defect, "measured" by this method, is actually smaller than the true one, i.e Alexander polynomial is unit.
As suggested in sec.\ref{3pretz}, we illustrate this with the example of triple pretzels.
%It is instructive to comment on the relation with "alternative definition" of defect in the
%sec.\ref{defAlex} -- to better illustrate and understand its limitation.
More sophisticated examples will be provided in sec.\ref{compexa} below.

For Alexander description the only relevant one is the fundamental representation, where
\be
{\cal F}_{[1]}^{(2n-1,2m-1,2l-1) }(A) = -\frac{A^{2(n+m+l-1)}+A^{2(n+m+l-2)} - A^{2(n+m-1)}-A^{2(n+l-1)}-A^{2(m+l-1)} +1}{\{A\}^2}
\ee
so that
\be
{\rm Al}_{[1]}^{(2n-1,2m-1,2l-1)} = 1 + (nm+nl+ml-m-n-l+1)\cdot\{q\}^2
\ee

Clearly, one can adjust $(n,m,l)$ so that the second term vanishes.
For example, it does so for $(N,M,L)=(-3,5,7)$, i.e. ${\bf F}_{[1]}^{(-3,5,7)}\sim\{A\}$.
However, no extra factorization occurs for other ${\bf F}_{[s]}^{(-3,5,7)}$ with $s>1$,
and the rest of the pattern follows the standard one for defect zero:

\be
\begin{picture}(300,100)(-40,-60)
\put(0,-20){
\put(-90,0){\mbox{defect $\delta^{\cal K}=0$:}}
\put(-100,0){
\put(180,-20){\line(1,0){75}}
\put(180,-10){\line(1,0){75}}
\put(200,0){\line(1,0){55}}
\put(210,10){\line(1,0){45}}
\put(220,20){\line(1,0){35}}
\put(230,30){\line(1,0){25}}
\put(240,40){\line(1,0){15}}
\put(250,50){\line(1,0){5}}
\put(180,-20){\line(0,1){10}}
\put(190,-20){\line(0,1){10}}
\put(200,-20){\line(0,1){20}}
\put(210,-20){\line(0,1){30}}
\put(220,-20){\line(0,1){40}}
\put(230,-20){\line(0,1){50}}
\put(240,-20){\line(0,1){60}}
\put(250,-20){\line(0,1){70}}
\put(160,-35){\mbox{$s$}}
\put(183,-35){\mbox{{\footnotesize $1$}}}
\put(193,-35){\mbox{{\footnotesize $2$}}}
\put(203,-35){\mbox{{\footnotesize $3$}}}
\put(213,-35){\mbox{{\footnotesize $4$}}}
\put(223,-35){\mbox{{\footnotesize $5$}}}
\put(233,-35){\mbox{{\footnotesize $6$}}}
\put(243,-35){\mbox{{\footnotesize $7$}}}
\put(253,-35){\mbox{{\footnotesize $8$}}}
%\put(263,-35){\mbox{{\footnotesize $9$}}}
%\put(270,-35){\mbox{{\footnotesize $10$}}}
%\put(280,-35){\mbox{{\footnotesize $11$}}}
%\put(290,-35){\mbox{{\footnotesize $12$}}}
%\put(300,-35){\mbox{{\footnotesize $13$}}}
%\put(310,-35){\mbox{{\footnotesize $14$}}}
%
\put(-50,0){
\put(320,-18){\mbox{{\footnotesize $1$}}}
\put(320,-8){\mbox{{\footnotesize $2$}}}
\put(320,2){\mbox{{\footnotesize $3$}}}
\put(320,12){\mbox{{\footnotesize $4$}}}
\put(320,22){\mbox{{\footnotesize $5$}}}
\put(320,32){\mbox{{\footnotesize $6$}}}
\put(320,42){\mbox{{\footnotesize $7$}}}
}
\put(170,31){\mbox{$\nu_s^{\cal K} = s-1 $}}
\put(180,-12.5){\line(1,0){10}}
\put(180,-15){\line(1,0){10}}
\put(180,-17.5){\line(1,0){10}}
}}
\label{degF1for3pret}
\end{picture}
\ee
In this sense the extra box at $s=1$ is just {\bf accidental}.
However, such {\it accidents} make the very convenient "definition" of defect
{\it a la} sec.\ref{defAlex} somewhat limited -- one can not fully trust it,
more thorough analysis, involving higher representations can be needed.
At the same time we see that a natural way to fight against
 the ambiguities with such definition of defect for particular knots
 is to consider entire families.

Coming bact to $(-3,5,7)$, note
that some additional simplifications occur for this knot
in higher representations:
\be
{\cal F}_{[s]}^{(-3,5,7)}(A=1)\sim\{q\} \ \ \ \ {\rm for} \ \ s\geq 2
\ee
However, this is not interesting --
it is a direct corollary of factorization property of the
{\it special} polynomial at $q=1$ \cite{DMMSS}:
\be
H_R(q=1,A) = \Big(1 + {\cal F}_{[1]}(q=1,A)\cdot \{A\}^2\Big)^{|R|}
\ee
-- then vanishing of ${\cal F}_{[1]}(A=1,q)$ implies that all
${\bf F}_{[s]}(A=1,q) \sim \{q\}$.

In fact, $(-3,5,7)$ is not a unique "accident" in the triple-pretzel family.
There are many other solutions
to the constraint  $l=-\frac{(m-1)(n-1)}{m+n-1} \in \mathbb{Z}$.
e.g. $(-5,9,11), \ (-5,7,17),\ (-7,9,31), \ \ldots$
However, extra degenerations of ${\cal F}_{[s]}$ with $s\geq 2$ never occur
in this series of accidental zeroes of ${\cal F}_{[1]}(A=1)$ --
the pattern of DE is always like (\ref{degF1for3pret}).\footnote{
There is a single exception -- the entire series $(-1,1,2l-1)$
consists of unknots and all $F_{[s]}^{(-1,1,2l-1)}=0$}
In fact, this is not a big surprise, because, say, the degeneration condition
${\cal F}_{[1]}(A=q^{-1})$ is a non-trivial Laurent polynomial in $q$,
and all of its coefficients do not vanish simultaneously for any
triple of variables $n,m,l$.

Still, for some other series of knots such improbable multiple degenerations can happen, as we will see in the next section.

In search for anomalies we checked the defects for all knots up to 11 intersections with {\it quadratic} Alexander polynomials. The knot data was taken from wonderful source \cite{knotinfo} and collected in Table \ref{defect0 data}.
\begin{table}[h!]
    \centering
    \begin{doublespace}
    \begin{tabular}{|c|c|c|c||c|c|c|c|}
     \hline
       Knot ${\cal K}$ &  Pretzel not. & $\delta^{{\cal K}}$ & $\text{Al}^{{\cal K}}$ & Knot ${\cal K}$ &  Pretzel not. & $\delta^{{\cal K}}$ & $\text{Al}^{{\cal K}}$   \\
     \hline
     \hline
     $3_1$ & $\left( \overline{1}, \overline{1}, \overline{1}\right)$ & 0 & $1 + \{q \}^2$& $9_{46}$ & $\left(- \overline{3}, \overline{3}, \overline{3}\right)$ & 0 & $1 - 2\{q \}^2$  \\
     \hline
     $4_1$ & $\left( -\overline{3}, \overline{1}, \overline{1}\right)$ & 0 & $1 - \{q \}^2$ &$10_{1}$ & $\left(- \overline{9}, \overline{1}, \overline{1}\right)$ & 0 & $1 - 4\{q \}^2$ \\
     \hline
     $5_2$ & $\left( \overline{3}, \overline{1}, \overline{1}\right)$ & 0 & $1 + 2\{q \}^2$ & $10_{3}$ & $\left(- \overline{7}, \overline{3}, \overline{1}\right)$ & 0 & $1 - 6\{q \}^2$ \\
     \hline
     $6_1$ & $\left( -\overline{5}, \overline{1}, \overline{1}\right)$ & 0 & $1 -2 \{q \}^2$ & $11a247$ & $\left( \overline{9}, \overline{1}, \overline{1}\right)$ & 0 & $1 + 5\{q \}^2$ \\
     \hline
     $7_2$ & $\left( \overline{5}, \overline{1}, \overline{1}\right)$ & 0 &$1 + 3\{q \}^2$ & $11a343$ & $\left( \overline{7}, \overline{3}, \overline{1}\right)$ & 0 & $1 + 8\{q \}^2$  \\
     \hline
     $7_4$ & $\left( \overline{3}, \overline{3}, \overline{1}\right)$ & 0 & $1 + 4\{q \}^2$& $11a362$ & $\left( \overline{5}, \overline{3}, \overline{3}\right)$ & 0 & $1 + 10\{q \}^2$ \\
     \hline
     $8_1$ & $\left( -\overline{7}, \overline{1}, \overline{1}\right)$ & 0 &$1 -3 \{q \}^2$ & $11a363$ & $\left( \overline{5}, \overline{5}, \overline{1}\right)$ & 0 & $1 + 9\{q \}^2$ \\
     \hline
     $8_3$ & $\left( -\overline{5}, \overline{3}, \overline{1}\right)$ & 0 &$1 -4 \{q \}^2$ & $11n67$ & not a 3-pretz. & 0 & $1 -2\{q \}^2$ \\
     \hline
     $9_2$ & $\left( \overline{7}, \overline{1}, \overline{1}\right)$ & 0 & $1 + 4\{q \}^2$&  $11n97$ & not a 3-pretz. & 0 & $1 -2\{q \}^2$ \\
     \hline
     $9_5$ & $\left( \overline{5}, \overline{3}, \overline{1}\right)$ & 0 & $1 + 6\{q \}^2$& $11n139$ & $\left( \overline{5}, \overline{3}, -\overline{3}\right)$ & 0 & $1 -2\{q \}^2$  \\
     \hline
     $9_{35}$ & $\left( \overline{3}, \overline{3}, \overline{3}\right)$ & 0 & $1 + 7\{q \}^2$&$11n141$ & $\left( -\overline{5}, \overline{3}, \overline{3}\right)$ & 0 & $1 - 5\{q \}^2$  \\
     \hline
    \end{tabular}
    \end{doublespace}
    \caption{The table provides data for knots with quadratic Alexander polynomials.}
    \label{defect0 data}
\end{table}

\section{More complicated examples
\label{compexa}}

\subsection{$\left(N,M,\overline{K}\right)$ pretzel knots}
The one-parametric series $(3,3,\overline{2k})$ with two parallel and one antiparallel fingers
is described by
\be
H_R^{(3,3,\overline{2k})} = d_R^2\sum_{X\in R\otimes \bar R}
\frac{{\cal A}^{\rm{par}}_{1X}(3)\,{\cal A}^{\rm{par}}_{1X}(3) \,
{\cal A}^{\rm{ea}}_{1X}(\overline{2k})}{\sqrt{d_X}}
\ee
has defect $\delta^{(3,3,\overline{2k})}=2$ for all values of $k$, except $k=0$.
This illustrates the {\bf antiparallel invariance of defect}.

For particular value of $k=0$ this series contains a composite of two trefoils.
%and the defect drops down to $\delta^{(3,3,0)}=1$.
%Note that
Defect of a composites differs significantly from that of the constituents.
Explicit expression is easily obtained from the Alexander property of the defect.
Since the reduced HOMFLY-PT is a product of two constituent HOMFLY-PT,
the same is true for Alexander polynomials, and
\be
{\rm Al}^{{\cal K}_1\cup{\cal K}_2} = {\rm Al}^{{\cal K}_1} \cdot {\rm Al}^{{\cal K}_2}
\ \stackrel{(\ref{Aldef})}{=} \
\left(1 + \{q\}^2 \cdot {\rm Pol}^{{\cal K}_{\delta_1} } \right)
\left(1 + \{q\}^2 \cdot {\rm Pol}^{{\cal K}_{\delta_2} } \right)
=  \nn \\ =
1 + \{q\}^2 \cdot
\left({\rm Pol}^{{\cal K}_{\delta_1}} + {\rm Pol}^{{\cal K}_{\delta_2} }
 + \{q\}^2 \cdot  {\rm Pol}^{{\cal K}_{\delta_1} } \cdot {\rm Pol}^{{\cal K}_{\delta_2} }  \right)
 \ \ \ \ \ \  \ \ \ \ \ \
\ee
implies that
\be
\delta^{{\cal K}_1\cup{\cal K}_2}  =\delta^{{\cal K}_1}+ \delta^{{\cal K}_2}+  1
\label{compositedefect}
\ee
For our particular example of $(3,3,0)$ this means that
$\delta^{(3,3,0)}=1$, i.e. the defect {\it accidentally} drops down from 2 to 1
in a particular member $k=0$ of the family.
%This is confirmed by explicit formula
%\be
%{\rm Al}_{[1]}^{(2n-1,2m-1,\overline{2k})} = 1 + ??? ???
%\ee

\bigskip

For the full family $(2n-1,2m-1,\overline{2k})$
\be
H_R^{(2n-1,2m-1,\overline{2k})} = d_R^2 \sum_{X\in R\otimes \bar R}
\frac{  {\cal A}^{\rm{par}}_{1X}(2n-1) \, {\cal A}^{\rm{par}}_{1X}(2m-1) \, {\cal A}^{\rm{ea}}_{1X}(\overline{2k}) }{\sqrt{d_X}}
\ee
the defect is
\be
\delta^{(2n-1,2m-1,\overline{2k})} = n+m-2
%= \frac{(2n-1)+(2m-1)-2}{2}
\ \ \ {\rm for} \ \ n,m\geq 1
\label{posPPAdefect}
\ee
i.e. {\bf does not depend on the antiparallel evolution and depends linearly on the
parallel ones}.
This is in full accordance with our general suggestion.

Again, at $k=0$ we get a composite of the 2-strand torus knots $(2n-1,1)$ and $(2m-1,1)$
and from (\ref{compositedefect})  and (\ref{2torusdefect})
the defect is $n+m-3$ (for $n,m\geq 1$),
%$\frac{|2n-1|+|2m-1|-4}{2}$,
i.e. at $k=0$ there is an accidental decrease of the defect by one.
In the particular case $(1,1,\overline{2k})$ at $k=0$ the composite of two
unknots is an unknot, and our formula is consistent with $\delta^{\rm unknot}=-1$,
suggested in (\ref{unknotdefect}).

\bigskip

If  $n$ or $m$ are not positive,  the formula (\ref{posPPAdefect}) gets slightly more involved:
\be
\delta^{(2n-1,2m-1,\overline{2k})} = \left|n-\frac{1}{2}\right| + \left|m-\frac{1}{2}\right|
+ \frac{{\rm sign}\left(n-\frac{1}{2}\right)\cdot{\rm sign}\left(m-\frac{1}{2}\right)}{2}
-\frac{3}{2}
% - \underline{\delta_{k,0}}
%\ \ \ \ \ \ {\rm for} \ \ k\neq 0
\ee
It is apparently symmetric under the permutation of $n$ and $m$ and under the
simultaneous sign change of $2n-1$ and $2m-1$.
This expression is obtained from both criteria (\ref{extrafactor}) and (\ref{Aldef}) and
demonstrates the relatively sophisticated structure of the switching region
between positive and negative parameters of the parallel evolution.

The drop of defect by one takes place at $k=0$ when $2n-1$ and $2m-1$ have the same sign.
Otherwise this happens at $2k= 2$ when $2n-1>0$ and $2m-1=- 1$ or $2n-1=-1$ and $2m-1>0$
and at $2k=-2$ when $2n-1<0$ and $2m-1=1$ or $2n-1=1$ and $2m-1<0$. In the regions of different signes of $2n-1$ and $2m-1$ no drop of the defect is observed. Also there are unknots when $2n-1=-(2m-1)=\pm 1$.

\begin{table}[h!]
\centering
    \begin{tabular}{|c||c|c|c|c|c|c|c|c|c|c|}
    \hline
    \diagbox{\raisebox{0.1cm}{{\footnotesize $2m-1$}}}{\raisebox{-0.1cm}{{\footnotesize $ 2n-1$}}}
    & -7 & -5 & -3 & -1 & 1 & 3 & 5 & 7   \\
    \hline
    \hline
    -7 & 6 & 5 & 4 & \textcolor{blue}{3} & \textcolor{red}{2} & \textcolor{blue}{3} & 4 & 5   \\
    \hline
    -5 & 5 & 4 & \textcolor{blue}{3} & \textcolor{red}{2} & \textcolor{violet}{1} & \textcolor{red}{2} & \textcolor{blue}{3} & 4   \\
    \hline
    -3 & 4 & \textcolor{blue}{3} & \textcolor{red}{2} & \textcolor{violet}{1} & \textcolor{green}{0} & \textcolor{violet}{1} & \textcolor{red}{2} & \textcolor{blue}{3}    \\
    \hline
    -1 & \textcolor{blue}{3} & \textcolor{red}{2} &\textcolor{violet}{1} & \textcolor{green}{0} & $\emptyset$ & \textcolor{green}{0} & \textcolor{violet}{1} & \textcolor{red}{2}   \\
    \hline
     1& \textcolor{red}{2} & \textcolor{violet}{1} & \textcolor{green}{0} & $\emptyset$ & \textcolor{green}{0} & \textcolor{violet}{1} & \textcolor{red}{2} & \textcolor{blue}{3}  \\
    \hline
     3& \textcolor{blue}{3} &  \textcolor{red}{2} & \textcolor{violet}{1} &\textcolor{green}{0} & \textcolor{violet}{1} & \textcolor{red}{2} & \textcolor{blue}{3} & 4   \\
    \hline
     5& 4 & \textcolor{blue}{3} & \textcolor{red}{2} & \textcolor{violet}{1} & \textcolor{red}{2} & \textcolor{blue}{3} & 4 & 5   \\
    \hline
    7 & 5 & 4 & \textcolor{blue}{3} & \textcolor{red}{2} & \textcolor{blue}{3} & 4 & 5 & 6  \\
    \hline
\end{tabular}
\hspace{0mm}
    \begin{tabular}{|c||c|c|c|c|c|c|c|c|c|c|}
    \hline
    \diagbox{\raisebox{0.1cm}{{\footnotesize $2m-1$}}}{\raisebox{-0.1cm}{{\footnotesize $2n-1$}}}
    & -7 & -5 & -3 & -1 & 1 & 3 & 5 & 7   \\
    \hline
    \hline
    -7 & 0 & 0 & 0 & 0 & \textcolor{red}{-1} &  &  &     \\
    \hline
    -5 & 0 & 0 & 0 & 0 & \textcolor{red}{-1} &  &  &     \\
    \hline
    -3 & 0 & 0 & 0 & 0 & \textcolor{red}{-1} &  &  &   \\
    \hline
    -1 & 0 & 0 & 0 & 0 & $\emptyset$ & \textcolor{red}{1} & \textcolor{red}{1} & \textcolor{red}{1} \\
    \hline
    1 & \textcolor{red}{-1} & \textcolor{red}{-1} & \textcolor{red}{-1} & $\emptyset$ & 0 & 0 & 0 & 0   \\
    \hline
    3 &  &  &  & \textcolor{red}{1} & 0 & 0 & 0 & 0   \\
    \hline
    5 &  &  &  & \textcolor{red}{1} & 0 & 0 & 0 & 0   \\
    \hline
    7 &  &  &  & \textcolor{red}{1} & 0 & 0 & 0 & 0   \\
    \hline
\end{tabular}

 \caption{\footnotesize
 The left table lists the defects $\delta^{(2n-1,2m-1,\overline{2k})}$.
 The right table shows the critical values of $k$ where the defect decreases by one.
 The empty box means that the defect does {\it not} drop for these values of $n$ and $m$.
 The item $\emptyset$ means unknot.}
    \label{PPA defects}
\end{table}
The left Table \ref{PPA defects} is symmetric with respect to the main diagonal, since on the topological level pretzel knots $(2n-1,2m-1,\overline{2k})$ and $(2m-1,2n-1,\overline{2k})$ are identical. Moreover, this table is symmetric with respect to the {\it secondary} diagonal. This fact follows from the relation:
\begin{equation}
    \delta^{ (2n-1,2m-1,\overline{2k})} = \delta^{(-2n+1,-2m+1,-\overline{2k})}
\end{equation}
That is a simple corollary of two facts:
\begin{enumerate}
    \item Mirror image of a knot $\bar{\mathcal{K}}$ has HOMFLY polynomial $H^{\bar{\mathcal{K}}}_{R}(A,q) = H^{\mathcal{K}}_{R}(A^{-1},q^{-1})$, while the reflection $A \to A^{-1}, q \to q^{-1}$ changes differentials $\{ A q^{k} \}$ only by sign. Therefore mirror image of a knot has the same defect.
    \item  $\delta^{(-2n+1,-2m+1,-\overline{2k})} = \delta^{(-2n+1,-2m+1,\overline{2k})}$ due to the hypothesis of the anti-parallel evolution
\end{enumerate}

\subsection{KTC  mutant and its relatives}

%Counterexample:  knots with defect -1  (unit Alexander). ???

%???

%${\rm Al}^{{\cal K}_0}_{[1]} = 1 + {\rm const}\cdot \{q\}^2$
%while for $\delta^{\cal K}=-1$ it is just unity:
%\be
%{\rm Al}^{{\cal K}_{-1}}_{[1]} = 1
%\ee
%The first example of such knot appears at 11??? intersection, and it is the celebrated
%KTC??? mutant.

KTC mutants $11n34 \ \& \ 11n42$ are $(3, -2|\bar 2| -3, 2)\ \& \ (3, -2|\bar 2|2, -3)$ have unit Alexander and provide the most anomalous examples of DE structure. In the fundamental representation
\be
H_{[1]}^{[11n34]} =  H_{[1]}^{[11n42]}:=
1 - \frac{[6][4]}{[2]^2[3]A} \{Aq\}\underline{\{A\}}\{A/q\}
\ \ \ \ \ \Longrightarrow \ \ \ \ \
{\rm Al}_{[1]}^{[11n34]} = 1
\ee
there is one extra factor, $\{A\}$, in $F_{[1]}^{[11n34]}$.

The difference between the two mutants shows up already for
the first non-rectangular representation $H_{[21]}$,
but it remains absent for arbitrary large symmetric and even rectangular ones.
In this paper we are not so much interested in {\it mutant} property \cite{morton2009mutant,Morton,Bishler2020, Bishler:2020wbq},
our emphasize here is rather on the trivial Alexander
(we just use the well know fact that it is unity for the KTC pair).
Therefore we concentrate on symmetric representations.
Due to the property \cite{DMMSS} ${\rm Al}_{[r]}(q) = {\rm Al}_{[1]}(q^r)$
(actually true for all one-hook representations)
Alexander will be remain unity for all of them.\footnote{
It deserves noting that for $\delta>0$ this is a highly non-trivial property,
since in this case the higher coefficients of the DE contribute to Alexander.
For example, for the defect-3 torus knot ${\rm Torus}[3,4]=8_{19}$
the first symmetric Alexander
${\rm Al}_{[2]}^{8_{19}} = 1 - F_{[1]}^{8_{19}}(A=1,q)\cdot \{q\}^2 -
{\bf F}_{[2]}^{8_{19}}(A=1,q)\cdot\{q^3\}\{q^2\}\{q\}$,
and the last term artfully compensates for the difference between
$F_{[1]}(A=1,q)$ and the needed $F_{[1]}(A=1,q^2)$.
}

KTC mutants are arborescent knots of type $(p,q | \overline{2k} | s,r)$, but not pretzels,
they contain a "propagator" between two double-finger vertices
and  are described by (57) of \cite{MMMRS}:

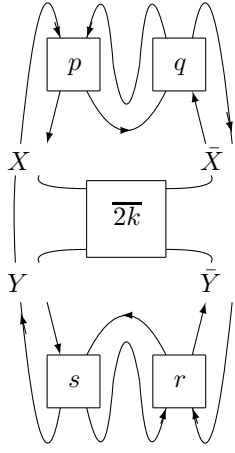
\be
\begin{picture}(300,170)(-130,-135)
\put(0,0){\line(1,0){20}}
\put(0,0){\line(0,1){20}}
\put(0,20){\line(1,0){20}}
\put(20,0){\line(0,1){20}}
\put(8,8){\mbox{$p$}}
\put(40,0){\line(1,0){20}}
\put(40,0){\line(0,1){20}}
\put(40,20){\line(1,0){20}}
\put(60,0){\line(0,1){20}}
\put(48,8){\mbox{$q$}}
\qbezier(15,20)(22,50)(25,10)
\qbezier(45,20)(38,50)(35,10)
\qbezier(25,10)(30,-20)(35,10)
\qbezier(15,0)(30,-30)(45,0)
\qbezier(5,20)(-2,60)(-10,-20)
\qbezier(55,20)(62,60)(70,-20)
\put(5,0){\vector(-1,-4){5}}
\put(60,-20){\vector(-1,4){5}}
%
%\put(30,-4){\vector(-1,0){2}}
\put(17,26){\vector(-1,-3){2}}
\put(30,-15){\vector(1,0){2}}
\put(3,26){\vector(1,-3){2}}
\put(68,-8){\vector(1,-4){2}}
\put(-15,-30){\mbox{$X$}}
\put(58,-30){\mbox{$\bar X$}}
\put(0,-100){\line(1,0){20}}
\put(0,-100){\line(0,-1){20}}
\put(0,-120){\line(1,0){20}}
\put(20,-100){\line(0,-1){20}}
\put(8,-112){\mbox{$s$}}
\put(40,-100){\line(1,0){20}}
\put(40,-100){\line(0,-1){20}}
\put(40,-120){\line(1,0){20}}
\put(60,-100){\line(0,-1){20}}
\put(48,-112){\mbox{$r$}}
\qbezier(15,-120)(22,-150)(25,-110)
\qbezier(45,-120)(38,-150)(35,-110)
\qbezier(25,-110)(30,-80)(35,-110)
\qbezier(15,-100)(30,-70)(45,-100)
\qbezier(5,-120)(-2,-160)(-10,-80)
\qbezier(55,-120)(62,-160)(70,-80)
\put(55,-100){\vector(1,4){5}}
\put(0,-80){\vector(1,-4){5}}
%
%\put(30,-4){\vector(-1,0){2}}
\put(43,-126){\vector(1,3){2}}
\put(30,-85){\vector(-1,0){2}}
%
%\put(-9.5,-15){\vector(1,4){2}}
\put(57,-126){\vector(-1,3){2}}
\put(-8,-92){\vector(-1,4){2}}
\put(-15,-76){\mbox{$Y$}}
\put(58,-76){\mbox{$\bar Y$}}
\qbezier(-12,-32)(-13,-45)(-12,-65)
\qbezier(71.5,-32)(73,-45)(71.5,-65)
\qbezier(-3,-32)(-6,-37)(15,-37)
\qbezier(-3,-65)(-6,-60)(15,-60)
\qbezier(62,-32)(65,-37)(45,-37)
\qbezier(62,-65)(65,-60)(45,-60)
\put(15,-34){\line(0,-1){29}}  \put(45,-34){\line(0,-1){29}}
\put(15,-34){\line(1,0){30}}  \put(15,-63){\line(1,0){30}}
\put(25,-51){\mbox{$\overline{2k}$}}
\end{picture}
\label{mutfam}
\ee
Therefore, the reduced HOMFLY polynomial for the diagram (\ref{mutfam}) is
\be
%\sum_{X,Y\in R\otimes \bar R} \frac{d_R^4}{\sqrt{d_Xd_Y}} \overline{\sum_{a,b,c=1}^{m_X}}
% \overline{\sum_{d,e,f=1}^{m_Y}}
%{\cal A}^{\rm par}(p)_{1X}^{ab} {\cal A}^{\rm par}(q)_{1X}^{bc }
% \left(\sum_{Z \in R\otimes \bR}\overline{\sum_{g,h,i}^{m_Z}}
% \frac{1}{\sqrt{d_Z}}\bar S_{XZ}^{ca,gh}
%{\cal A}^{\rm ea}(m)_{1Z}^{hi }\bar S_{YZ}^{ig,fd}\right)
%{\cal A}^{\rm par}(r)_{1Y}^{de} {\cal A}^{\rm par}(s)_{1Y}^{ef }
%= \nn \\ =
H_{R}^{(p,q|\overline{2k}|s,r)}= d_R^2 \sum_{Z\in R\otimes \bar R} \frac{ {\cal A}^{ppS}_Z(p,q)\,
{\cal A}^{\rm ea}_{1Z}(\overline{2k})\,
 {\cal A}^{ppS}_Z(r,s) }{\sqrt{d_Z}}
\label{Mpqmrs}
\ee
where
\be
{\cal A}^{ppS}_Z(N,M) =
d_{R} \sum_{X\in R\otimes \bar R} \frac{
{\cal A}^{\rm par}_{1X}(N) \, \bar S_{XZ} \, {\cal A}^{\rm par}_{1X}(M)
 }{\sqrt{d_X}}
\label{defAppS}
\ee
and expressions ${\cal A}$ for particular fingers are provided in sec.\ref{prefing}.
The four external fingers involve parallel braids, while that of the even length $2k$ in the propagator
is antiparallel. The evolution along this obvious antiparallel braid from the KTC mutant generates a family
$(3,-2|\overline{2k}|-3,2)$
with the following structure of the DE coefficients:
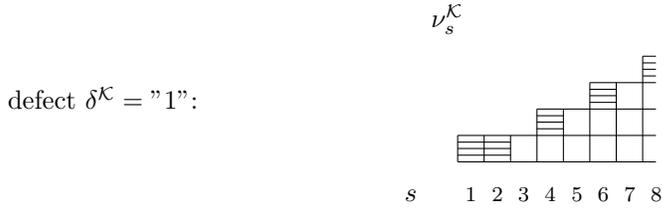
\be
\begin{picture}(300,100)(-40,-60)
\put(0,-20){
\put(-90,0){\mbox{defect $\delta^{\cal K}="1"$:}}
\put(-100,0){
\put(180,-20){\line(1,0){75}}
\put(180,-10){\line(1,0){75}}
%\put(200,0){\line(1,0){55}}
\put(210,0){\line(1,0){45}}
\put(230,10){\line(1,0){25}}
\put(250,20){\line(1,0){5}}
%\put(240,40){\line(1,0){15}}
%\put(250,50){\line(1,0){5}}
%
\put(180,-20){\line(0,1){10}}
\put(190,-20){\line(0,1){10}}
\put(200,-20){\line(0,1){10}}
\put(210,-20){\line(0,1){20}}
\put(220,-20){\line(0,1){20}}
\put(230,-20){\line(0,1){30}}
\put(240,-20){\line(0,1){30}}
\put(250,-20){\line(0,1){40}}
\put(160,-35){\mbox{$s$}}
\put(183,-35){\mbox{{\footnotesize $1$}}}
\put(193,-35){\mbox{{\footnotesize $2$}}}
\put(203,-35){\mbox{{\footnotesize $3$}}}
\put(213,-35){\mbox{{\footnotesize $4$}}}
\put(223,-35){\mbox{{\footnotesize $5$}}}
\put(233,-35){\mbox{{\footnotesize $6$}}}
\put(243,-35){\mbox{{\footnotesize $7$}}}
\put(253,-35){\mbox{{\footnotesize $8$}}}
%\put(263,-35){\mbox{{\footnotesize $9$}}}
%\put(270,-35){\mbox{{\footnotesize $10$}}}
%\put(280,-35){\mbox{{\footnotesize $11$}}}
%\put(290,-35){\mbox{{\footnotesize $12$}}}
%\put(300,-35){\mbox{{\footnotesize $13$}}}
%\put(310,-35){\mbox{{\footnotesize $14$}}}
%
\put(-50,0){
\put(320,-18){\mbox{{\footnotesize $1$}}}
\put(320,-8){\mbox{{\footnotesize $2$}}}
\put(320,2){\mbox{{\footnotesize $3$}}}
\put(320,12){\mbox{{\footnotesize $4$}}}
%\put(320,22){\mbox{{\footnotesize $5$}}}
%\put(320,32){\mbox{{\footnotesize $6$}}}
%\put(320,42){\mbox{{\footnotesize $7$}}}
%
}
\put(170,31){\mbox{$\nu_s^{\cal K}$}}
\put(180,-12.5){\line(1,0){20}}
\put(180,-15){\line(1,0){20}}
\put(180,-17.5){\line(1,0){20}}
\put(210,-2.5){\line(1,0){10}}
\put(210,-5){\line(1,0){10}}
\put(210,-7.5){\line(1,0){10}}
\put(230,7.5){\line(1,0){10}}
\put(230, 5){\line(1,0){10}}
\put(230,2.5){\line(1,0){10}}
\put(250,17.5){\line(1,0){5}}
\put(250,15){\line(1,0){5}}
\put(250,12.5){\line(1,0){5}}
}}
\end{picture}
\ee
Again, the extra boxes at are just accidental,
still now it is an "accident" which appears
simultaneously at many places
and for the whole one-parametric {\it family} $(3,-2|\overline{2k}|-3,2)$. If one performs antiparallel evolution at any crossing except antiparallel braid $\overline{2k}$, all the extra boxes {\it vanish} and the defect becomes $\delta^{{\cal K}} = 1$. Therefore we treat black boxes as "accident", appearing in a highly symmetric point in the space of knots - KTC mutant.

%This is not??? the pattern for defect 1???.
%???

%???Moreover, the coefficients also have simple evolution properties.???

%\subsection{A ladder for higher defects $\delta^{\cal K}>1$}

\subsection{The family $(2a-1,2b\,\big|\,\overline{2k}\,\big|\,2c-1,2d)$}

Entire two-parametric families
\be
(2n+1,-2n\,\big|\,\overline{2k}\,\big|\,-2n-1,2n)
\ \ {\rm and} \ \
(2n-1,-2n\,\big|\,\overline{2k}\,\big|\,2n,-2n+1)
\ee
have unit fundamental Alexander, and $k$-independent defects.
However, these defects increase with $|n|$ and the structure of additional degenerations remain obscure.

Instead one can study the "plateaux" in parallel evolution
in the full 5-parametric family
$(2a-1,2b\,\big|\,\overline{2k}\,\big|\,2c-1,2d)$.

The family $(2a-1,2b\,\big|\,\overline{2k}\,\big|\,2c-1,2d)$
is symmetric under the permutations of fingers
\be
(2a-1,2b\,\big|\,\overline{2k}\,\big|\,2c-1,2d)
= (2b,2a-1\,\big|\,\overline{2k}\,\big|\,2c-1,2d)
=(2c-1,2d\,\big|\,\overline{2k}\,\big|\,2a-1,2b)
\ee
but one is not allowed to break the pairs:
\be
(2a-1,2b\,\big|\,\overline{2k}\,\big|\,2c-1,2d)\neq
(2a-1,2d\,\big|\,\overline{2k}\,\big|\,2c-1,2b)
\ \ {\rm if} \ \  a\neq c \ \ {\rm and} \ \ b\neq d
\ee
These symmetries are  respected by explicit formula for
the defect:
\be
\delta^{(2a-1,2b\,\big|\,\overline{2k}\,\big|\,2c-1,2d)}
= \left|a -\frac{1}{2}\right| + \left|c -\frac{1}{2}\right| + |b|+|d|-2\, +
%\nn
\ee
{\footnotesize
\be
\!\!\!\!\!\!\!\!\!\!\!\!
+\ \left\{
\ \ \ \ \ \ \
\begin{array}{ccc}
 a,c\leq 0 &\ \ \ &
%  + \overbrace{\theta(-a)\theta(-c)}^{\theta_{a\leq 0}\theta_{c\leq 0}}
+  \theta(-a)\theta(-c)\cdot \Big\{-2\cdot{\rm min}(b,d)\cdot\theta(b-1)\theta(d-1)
%+  \nn \\
%&&   \ \ \ \ \ \ \ \ \ \ \ \ \ \ \
 + \big(2\cdot{\rm min}(b,d)+2\cdot{\rm max}(a,c)-1\big)\cdot
 \theta\big({\rm min}(b,d)+{\rm max}(a,c)-1\big)\Big\}
\nn \\
\!\!\!\!\!\!\!\!\!\!\!\!  a\leq 0,\ c\geq 1 &&
+ \theta(-a)\theta(c-1)\Big\{-2\cdot {\rm min}\big(b,|d|\big)\cdot\theta(b-1)\theta(-d) 
+ \ldots\Big\}
\nn \\
\!\!\!\!\!\!\!\!\!\!\!\!  a\geq 1,\ c\leq 0 &&
+ \theta(a-1)\theta(-c)\cdot\Big\{-2\cdot {\rm min}\big(|b|,d\big)\cdot\theta(-b)\theta(d-1) 
%- \big(2\cdot{\rm min}(a-1,|c|)-2\cdot{\rm min}(|b|,d)+1\big)\cdot
%\big(-{\rm min}(a-1,|c|)+{\rm min)(|b|,d)\big)\cdot\theta(d-a)
+ \ldots \Big\}
\nn \\
 a,c\geq 1 &&
+ \theta(a-1)\theta(c-1)\Big\{
\underbrace{2\cdot{\rm max}(b,d)}_{-2\cdot {\rm min}\big(|b|,|d|\big)}\cdot\theta(-b)\theta(-d)
%-  \nn \\
%&&   \ \ \ \ \ \ \ \ \ \ \ \ \ \ \
- \big(2\cdot{\rm max}(b,d)+2\cdot{\rm min}(a,c)-1 \big)
\cdot\theta\big(-{\rm max}(b,d)-{\rm min}(a,c)\big)\Big\}
\nn
\end{array}\right.
\ee
}

\noindent
Correction terms in the second line do not grow when absolute value of any of the four variables 
becomes much bigger that the others -- the linear growth is provided by the main terms
in the first line.

\subsection{(Anti)parallel evolution for any arborescent knots}
Turns out, that the (anti)parallel evolution {\it does not} lead us out of the family of arborescent knots. For example, we provide picture of antiparallel evolution of a crossing in a parallel braid:
\be
\begin{picture}(300,120)(-50,-60)
\put(-50,-10){\line(1,0){40}}
\put(-50,-10){\line(0,1){20}}
\put(-50,10){\line(1,0){40}}
\put(-10,-10){\line(0,1){20}}
\qbezier(-70,-40)(-70,-5)(-50,-5)
\qbezier(-70,40)(-70,5)(-50,5)
\qbezier(10,-40)(10,-5)(-10,-5)
\qbezier(10,40)(10,5)(-10,5)
\put(-80,-40){\line(0,1){80}}
\put(20,-40){\line(0,1){80}}
\put(-85,-50){\mbox{$X$}}
\put(5,-50){\mbox{$\bar X$}}
\put(-85,43){\mbox{$Y$}}
\put(5,43){\mbox{$\bar Y$}}
\put(-45,-2){\mbox{$\overline{2m+1}$}}
\put(150,-10){\line(1,0){40}}
\put(150,-10){\line(0,1){20}}
\put(150,10){\line(1,0){40}}
\put(190,-10){\line(0,1){20}}
\qbezier(130,-50)(130,-5)(150,-5)
\qbezier(130,50)(130,5)(150,5)
\qbezier(210,-50)(210,-5)(190,-5)
\qbezier(210,50)(210,5)(190,5)
%
%\put(120,-40){\line(0,1){80}}
\put(220,-50){\line(0,1){100}}
\put(115,-60){\mbox{$X$}}
\put(205,-60){\mbox{$\bar X$}}
\put(115,53){\mbox{$Y$}}
\put(205,53){\mbox{$\bar Y$}}
\put(155,-2){\mbox{$\overline{2m+1}$}}
\put(130,-45.5){\circle*{4}}
\put(130,45.5){\circle*{4}}
\put(207,-21){\circle*{4}}
\put(207,21){\circle*{4}}
\put(65,-2){\mbox{$=$}}
\qbezier(120,-50)(120,-45)(165,-40)\qbezier(165,-40)(212,-40)(212,0)
\qbezier(120,50)(120,45)(165,40)\qbezier(165,40)(212,40)(212,0)
\qbezier[70](100,15)(150,15)(245,15)
\qbezier[70](100,-15)(150,-15)(245,-15)
\put(240,-25){\mbox{$Z$}}
\put(240,18){\mbox{$Z$}}
\end{picture}
\label{horbraid1}
\ee
For $m = 0$ the left picture is a part of the parallel braid, while for the other values of $m$ the horizontal antiparallel braid considered as the pretzel finger in the right picture. 
This is a contraction of three blocks, where the middle one is exactly the odd antiparallel pretzel finger from sec.\ref{prefing}.
%(\ref{elehor}).
Schematically, it is
\be
d_{R}^2\sum_{Z\in R\otimes R}  \frac{\Big(\bar{T} \bar{S} \bar{T} S\Big)_{YZ} \, {\cal A}^{\rm oa}\left(\overline{2m+1}\right)_{1Z} \, \Big(S^\dagger  \bar{T}^{-1} \bar{S} \bar{T}^{-1} \Big)_{ZX}
}{\sqrt{d_Z}}
\ee
Using this method we can check the defect evolution hypothesis for any arborescent knots. Similarly, one can write an answer for insertion of a parallel braid into a antiparallel one.

\subsection{1-loop family as an example of non-arborescent knots}
To check the defect evolution hypothesis out of the arborescent family we used the technique developed in \cite{MMMRS,MMMRSS}. The method allows to insert any arborescent propagator instead of any ${\cal R}$-matrix in a braid. This class of knots is called {\it 1-loop family} and it is richer than the arborescent knots. The simplest non-arborescent examples are provided by the 3-strand braid. For example, we provide simplest non-arborescent knot $8_{19}$ which is included in 8-parametric antiparallel family
$(\overline{n_0},\overline{n_1},\overline{n_2},\overline{n_3},\overline{n_4},\overline{n_5},\overline{n_6},\overline{n_7},) = (1,-1,1,-1,1,-1,1,-1)$ of defect $\delta = 2$.
\be
\begin{picture}(300,100)(20,-50)
\put(0,30){\vector(1,0){35}}
\put(0,0){\vector(1,0){35}}
\put(0,-30){\vector(1,0){35}}
\put(30,-30){\vector(1,0){45}}
\put(95,-30){\vector(1,0){60}}
\qbezier(35,30)(40,30)(40,25)
\qbezier(35,0)(40,0)(40,5)
\put(35,25){\line(1,0){20}}
\put(35,5){\line(0,1){20}}
\put(40,12){\mbox{$\overline{n_0}$}}
\put(35,5){\line(1,0){20}}
\put(55,5){\line(0,1){20}}
\qbezier(50,25)(50,30)(55,30)
\qbezier(50,5)(50,0)(55,0)
%
%\put(55,30){\vector(1,0){20}}
\put(55,0){\vector(1,0){20}}
\qbezier(75,0)(80,0)(80,-5)
\qbezier(75,-30)(80,-30)(80,-25)
\put(75,-5){\line(1,0){20}}
\put(75,-25){\line(0,1){20}}
\put(80,-18){\mbox{$\overline{n_1}$}}
\put(75,-25){\line(1,0){20}}
\put(95,-25){\line(0,1){20}}
\qbezier(90,-5)(90,0)(95,0)
\qbezier(90,-25)(90,-30)(95,-30)
\put(55,30){\vector(1,0){60}}
\put(95,0){\vector(1,0){20}}
\qbezier(115,30)(120,30)(120,25)
\qbezier(115,0)(120,0)(120,5)
\put(115,25){\line(1,0){20}}
\put(115,5){\line(0,1){20}}
\put(120,12){\mbox{$\overline{n_2}$}}
\put(115,5){\line(1,0){20}}
\put(135,5){\line(0,1){20}}
\qbezier(130,25)(130,30)(135,30)
\qbezier(130,5)(130,0)(135,0)
\put(135,30){\vector(1,0){20}}
\put(135,0){\vector(1,0){20}}
\qbezier(155,-30)(160,-30)(160,-25)
\qbezier(155,0)(160,0)(160,-5)
\put(155,-25){\line(1,0){20}}
\put(155,-5){\line(0,-1){20}}
\put(160,-17){\mbox{$\overline{n_3}$}}
\put(155,-5){\line(1,0){20}}
\put(175,-5){\line(0,-1){20}}
\qbezier(170,-25)(170,-30)(175,-30)
\qbezier(170,-5)(170,0)(175,0)
\put(155,30){\line(1,0){40}}
\put(175,-30){\vector(1,0){60}}
\put(175,0){\vector(1,0){20}}
\qbezier(195,30)(200,30)(200,25)
\qbezier(195,0)(200,0)(200,5)
\put(195,25){\line(1,0){20}}
\put(195,5){\line(0,1){20}}
\put(200,12){\mbox{$\overline{n_4}$}}
\put(195,5){\line(1,0){20}}
\put(215,5){\line(0,1){20}}
\qbezier(210,25)(210,30)(215,30)
\qbezier(210,5)(210,0)(215,0)
\put(215,30){\vector(1,0){60}}
\put(215,0){\vector(1,0){20}}
\qbezier(235,-30)(240,-30)(240,-25)
\qbezier(235,0)(240,0)(240,-5)
\put(235,-25){\line(1,0){20}}
\put(235,-5){\line(0,-1){20}}
\put(240,-17){\mbox{$\overline{n_5}$}}
\put(235,-5){\line(1,0){20}}
\put(255,-5){\line(0,-1){20}}
\qbezier(250,-25)(250,-30)(255,-30)
\qbezier(250,-5)(250,0)(255,0)
\put(255,-30){\vector(1,0){60}}
\put(255,0){\vector(1,0){20}}
%\put(235,30){\vector(1,0){20}}
%
\qbezier(275,30)(280,30)(280,25)
\qbezier(275,0)(280,0)(280,5)
\put(275,25){\line(1,0){20}}
\put(275,5){\line(0,1){20}}
\put(280,12){\mbox{$\overline{n_6}$}}
\put(275,5){\line(1,0){20}}
\put(295,5){\line(0,1){20}}
\qbezier(290,25)(290,30)(295,30)
\qbezier(290,5)(290,0)(295,0)
\put(295,30){\vector(1,0){60}}
\put(295,0){\vector(1,0){20}}
\qbezier(315,-30)(320,-30)(320,-25)
\qbezier(315,0)(320,0)(320,-5)
\put(315,-25){\line(1,0){20}}
\put(315,-5){\line(0,-1){20}}
\put(320,-17){\mbox{$\overline{n_7}$}}
\put(315,-5){\line(1,0){20}}
\put(335,-5){\line(0,-1){20}}
\qbezier(330,-25)(330,-30)(335,-30)
\qbezier(330,-5)(330,0)(335,0)
\put(335,-30){\vector(1,0){20}}
\put(335,0){\vector(1,0){20}}
\end{picture}
\ee

\section{Conclusion
\label{conc}}

This paper studies the {\it evolution} of the {\it defect} \cite{KM1}
of the differential (cyclotomic) expansion \cite{MMM2, IMMM, BM1,Morozov1, Morozov2,Morozov3,Morozov4, Morozov5, Morozov6, BJLMMMS,AMM, MMM1, KM1,KM2,KM3, Itoyama:2012qt,Itoyama:2012re,Habiro2007,Nawata:2015wya,KNTZ1,CLZ,Chen:2014fpa,Chen:2015rid,Chen:2015sol,Kawagoe:2012bt,Kawagoe:2021onh,BG, Gorsky:2013jxa,Gukov:2011ry,Dunfield:2005si, berest2021, lovejoy2019colored, lovejoy2017colored, hikami2015torus,garoufalidis2005analytic,garoufalidis2011asymptotics}.
Evolution arises when  one vertex of the knot diagram is substituted by a
2-strand braid
and describes the dependence of the answer (knot polynomial) on this exponent.
In the case of "locally build" polynomials, like HOMFLY-PT, this means that
one of the ${\cal R}$-matrices is raised to some power
and evolution describes the dependence of the answer (knot polynomials) on this exponent.
The power of a rank-four tensor can be defined in two different
ways, which are nicknamed parallel and antiparallel evolution:
$$
{\cal R}^{ia}_{kb}{\cal R}^{bj}_{al} \ \stackrel{{\rm apar}}{\longleftarrow} \
    {\cal R}^{ij}_{kl} \ \stackrel{{\rm par}}{\longrightarrow} \
    {\cal R}^{ij}_{ab}{\cal R}^{ab}_{kl}
$$
At the first glance this evolution has nothing to do with defect.
It is a local feature, while defect is a global characteristic.
It is easy to treat evolution as a property of Lagrangian (input in the QFT formalism),
while defect is the one  of the correlators (output).
Still these turn out to be intimately related.
This paper provides a lot of evidence for a very general conjecture:
invariance of defect under the antiparallel braid evolution.
Then, answering a naturally arising question about parallel evolution,
we find out that {\it it} changes the defect by one/step.
Altogether this looks like a strange and powerful statement,
which can attract more attention to the notion of know defects.
It adds to the mysteries of knot theory
and is a new manifestation of the still unrevealed symmetry (conspiracy)
of the non-perturbative calculations -- which can be significant far beyond
the $3d$ topological theories.

\bigskip

To summarize, what we now know  about the defect are two four claims from \cite{KM1}:

\bigskip

1) For any given knot ${\cal K}$ the DE coefficients ${\bf F}_Q^{\cal K}$
factorize for big enough representations $Q$,
the integer-valued defect $\delta^{\cal K}$ measures when and how this happens. \\

2) Factorization follows one of the {\it ladder} patterns, described in sec.\ref{ladders}.\\

3) Defect is related to the power of the fundamental Alexander in $q^{\pm 2}$.\\

4) Defect governs the stabilization property, mentioned in sec.\ref{stabil}.

\bigskip

In this paper we added three more facts:

\bigskip

5) {\bf Defect does not change under antiparallel evolution}.
%???
It can drop for particular lengths of the braid. \\

6) {\bf Defect changes linearly under parallel evolution}.
There is a non-trivial pattern of switching between negative and positive
evolution parameters. \\

7) When fundamental Alexander is unit (and probably in some other circumstances)
the patterns can be somewhat richer than in 2),
with added steps and certain irregularities.

\bigskip

The subtle details mentioned in 5)-7) do not contradict 2), but emphasize that
for particular knots (and even entire families) there
can be {\it additional} degeneracites in the coefficients of DE,
which are not captured by the defect itself.
%which imposes the imposes restrictions from below???.
The fully detailed structure of degeneracies is probably controlled by
a more sophisticated quantity,
which is not just a single number.

\section*{Acknowledgements}

We are indebted to E. Lanina, And.Morozov and A.Sleptsov for insights, conversations and help.

Our work was partly supported by the grants of the Foundation for the Advancement of Theoretical Physics
“BASIS” (A.M., N.T.), by the grant of Leonhard Euler International Mathematical Institute in Saint Petersburg № 075–15–2019–1619 (N.T.),  by the RFBR grant 20-01-00644 (N.T.) and by the joint RFBR grants 21-51-46010-CT\_a (A.M., N.T.), 21-52-52004-MOST (A.M.).

\printbibliography

@article{MST1,
    author = "Mishnyakov, V. and Sleptsov, A. and Tselousov, N.",
    title = "{A new symmetry of the colored Alexander polynomial}",
    eprint = "2001.10596",
    archivePrefix = "arXiv",
    primaryClass = "hep-th",
    doi = "10.1007/s00023-020-00980-8",
    journal = "Annales Henri Poincare",
    volume = "22",
    number = "4",
    pages = "1235--1265",
    year = "2021"
}

@article{MST2,
    author = "Mishnyakov, V. and Sleptsov, A. and Tselousov, N.",
    title = "{A Novel Symmetry of Colored HOMFLY Polynomials Coming from $\mathfrak {sl}(N|M)$ Superalgebras}",
    eprint = "2005.01188",
    archivePrefix = "arXiv",
    primaryClass = "hep-th",
    doi = "10.1007/s00220-021-04073-3",
    journal = "Commun. Math. Phys.",
    volume = "384",
    number = "2",
    pages = "955--969",
    year = "2021"
}

@article{BM1,
    author = "Bishler, L. and Morozov, A.",
    title = "{Perspectives of differential expansion}",
    eprint = "2006.01190",
    archivePrefix = "arXiv",
    primaryClass = "hep-th",
    doi = "10.1016/j.physletb.2020.135639",
    journal = "Phys. Lett. B",
    volume = "808",
    pages = "135639",
    year = "2020"
}

@article{Morozov1,
    author = "Morozov, A.",
    title = "{KNTZ trick from arborescent calculus and the structure of differential expansion}",
    eprint = "2001.10254",
    archivePrefix = "arXiv",
    primaryClass = "hep-th",
    doi = "10.1134/S0040577920080036",
    journal = "Theor. Math. Phys.",
    volume = "204",
    pages = "863--889",
    year = "2020"
}

@article{Morozov2,
    author = "Morozov, A.",
    title = "{Pentad and triangular structures behind the Racah matrices}",
    eprint = "1906.09971",
    archivePrefix = "arXiv",
    primaryClass = "hep-th",
    doi = "10.1140/epjp/s13360-020-00234-w",
    journal = "Eur. Phys. J. Plus",
    volume = "135",
    number = "2",
    pages = "196",
    year = "2020"
}

@article{Morozov3,
    author = "Morozov, A.",
    title = "{Extension of KNTZ trick to non-rectangular representations}",
    eprint = "1903.00259",
    archivePrefix = "arXiv",
    primaryClass = "hep-th",
    reportNumber = "ITEP/TH-04/19, IITP/TH-04/19",
    doi = "10.1016/j.physletb.2019.05.016",
    journal = "Phys. Lett. B",
    volume = "793",
    pages = "464--468",
    year = "2019"
}

@article{Morozov4,
    author = "Morozov, A.",
    title = "{On exclusive Racah matrices $\bar S$ for rectangular representations}",
    eprint = "1902.04140",
    archivePrefix = "arXiv",
    primaryClass = "hep-th",
    reportNumber = "ITEP/TH-02/19, IITP/TH-02/19, MIPT/TH-02/19",
    doi = "10.1016/j.physletb.2019.04.034",
    journal = "Phys. Lett. B",
    volume = "793",
    pages = "116--125",
    year = "2019"
}

@article{BJLMMMS,
    author = "Bai, C. and Jiang, J. and Liang, J. and Mironov, A. and Morozov, A. and Morozov, An. and Sleptsov, A.",
    title = "{Differential expansion for link polynomials}",
    eprint = "1709.09228",
    archivePrefix = "arXiv",
    primaryClass = "hep-th",
    reportNumber = "FIAN-TD-22-17, IITP-TH-26-17, ITEP-TH-16-17",
    doi = "10.1016/j.physletb.2018.01.026",
    journal = "Phys. Lett. B",
    volume = "778",
    pages = "197--206",
    year = "2018"
}

@article{MMM2,
    author = "Mironov, Andrei and Morozov, Alexei and Morozov, Andrey",
    title = "{On colored HOMFLY polynomials for twist knots}",
    eprint = "1408.3076",
    archivePrefix = "arXiv",
    primaryClass = "hep-th",
    reportNumber = "FIAN-TD-13-14, ITEP-TH-25-14",
    doi = "10.1142/S0217732314501831",
    journal = "Mod. Phys. Lett. A",
    volume = "29",
    number = "34",
    pages = "1450183",
    year = "2014"
}

@article{KM1,
    author = "Kononov, Ya. and Morozov, A.",
    title = "{On the defect and stability of differential expansion}",
    eprint = "1504.07146",
    archivePrefix = "arXiv",
    primaryClass = "hep-th",
    reportNumber = "ITEP-TH-9-15",
    doi = "10.1134/S0021364015120127",
    journal = "JETP Lett.",
    volume = "101",
    number = "12",
    pages = "831--834",
    year = "2015"
}

@article{Morozov5,
    author = "Morozov, A.",
    title = "{Differential expansion and rectangular HOMFLY for the figure eight knot}",
    eprint = "1605.09728",
    archivePrefix = "arXiv",
    primaryClass = "hep-th",
    reportNumber = "ITEP-TH-11-16, IITP-TH-09-16",
    doi = "10.1016/j.nuclphysb.2016.08.027",
    journal = "Nucl. Phys. B",
    volume = "911",
    pages = "582--605",
    year = "2016"
}

@article{Morozov6,
    author = "Morozov, A.",
    title = "{Factorization of differential expansion for antiparallel double-braid knots}",
    eprint = "1606.06015",
    archivePrefix = "arXiv",
    primaryClass = "hep-th",
    reportNumber = "IITP-TH-10-16",
    doi = "10.1007/JHEP09(2016)135",
    journal = "JHEP",
    volume = "09",
    pages = "135",
    year = "2016"
}

@article{KM2,
    author = "Kononov, Ya. and Morozov, A.",
    title = "{Rectangular superpolynomials for the figure-eight knot 4$_{1}$}",
    eprint = "1609.00143",
    archivePrefix = "arXiv",
    primaryClass = "hep-th",
    reportNumber = "ITEP-TH-21-16",
    doi = "10.1134/S0040577917110058",
    journal = "Theor. Math. Phys.",
    volume = "193",
    number = "2",
    pages = "1630--1646",
    year = "2017"
}

@article{IMMM,
    author = "Itoyama, H. and Mironov, A. and Morozov, A. and Morozov, An.",
    title = "{HOMFLY and superpolynomials for figure eight knot in all symmetric and antisymmetric representations}",
    eprint = "1203.5978",
    archivePrefix = "arXiv",
    primaryClass = "hep-th",
    reportNumber = "FIAN-TD-04-12, ITEP-TH-14-12, OCU-PHYS-364",
    doi = "10.1007/JHEP07(2012)131",
    journal = "JHEP",
    volume = "07",
    pages = "131",
    year = "2012"
}

@article{KM3,
    author = "Kononov, Ya. and Morozov, A.",
    title = "{On rectangular HOMFLY for twist knots}",
    eprint = "1610.04778",
    archivePrefix = "arXiv",
    primaryClass = "hep-th",
    reportNumber = "ITEP-TH-22-16, IITP-TH-16-16",
    doi = "10.1142/S0217732316502230",
    journal = "Mod. Phys. Lett. A",
    volume = "31",
    number = "38",
    pages = "1650223",
    year = "2016"
}

@article{KNTZ1,
    author = "M. Kameyama,  S.Nawata, R.Tao and H. D. Zhang",
    title = "{Cyclotomic expansions of HOMFLY-PT colored by rectangular Young diagrams}",
    eprint = "1902.02275",
    archivePrefix = "arXiv",
    primaryClass = "math.GT",
    doi = "10.1007/s11005-020-01318-5",
    journal = "Lett. Math. Phys.",
    volume = "110",
    pages = "2573--2583",
    year = "2020"
}

@article{CLZ,
    author = "Qingtao Chen, Kefeng Liu, Shengmao Zhu",
    title = "{Cyclotomic expansions for the colored HOMFLY-PT invariants of double twist knots}",
    eprint = "2110.03616",
    archivePrefix = "arXiv",
    primaryClass = "math.GT",
    year = "2021"
}

@misc{BG,
  eprint = "2101.08243",
  archivePrefix = "arXiv",
  author = {Beliakova, Anna and Gorsky, Eugene},
  title = {Cyclotomic expansions for $\mathfrak{gl}_N$ knot invariants via interpolation Macdonald polynomials},
  year = {2021},
}

@article{MMMRS,
    author = "Mironov, A. and Morozov, A. and Morozov, An. and Ramadevi, P. and Singh, Vivek Kumar",
    title = "{Colored HOMFLY polynomials of knots presented as double fat diagrams}",
    eprint = "1504.00371",
    archivePrefix = "arXiv",
    primaryClass = "hep-th",
    reportNumber = "FIAN-TD-01-15, IITP-TH-01-15, ITEP-TH-04-15, IITP-TH-02-15",
    doi = "10.1007/JHEP07(2015)109",
    journal = "JHEP",
    volume = "07",
    pages = "109",
    year = "2015"
}

@misc{knotinfo,
  url = "https://knotinfo.math.indiana.edu/",
}

@article{AMM,
    author = "Arthamonov, S. B. and Mironov, A. and Morozov, A.",
    title = "{Differential hierarchy and additional grading of knot polynomials}",
    eprint = "1306.5682",
    archivePrefix = "arXiv",
    primaryClass = "hep-th",
    reportNumber = "FIAN-TD-09-13, ITEP-TH-20-13",
    doi = "10.1007/s11232-014-0159-9",
    journal = "Theor. Math. Phys.",
    volume = "179",
    pages = "509--542",
    year = "2014"
}

@article{MMM1,
    author = "Mironov, A. and Morozov, A. and Morozov, An.",
    editor = "Ma, Wen-Xiu and Kaup, David",
    title = "{Evolution method and \textquotedblleft{}differential hierarchy\textquotedblright{} of colored knot polynomials}",
    eprint = "1306.3197",
    archivePrefix = "arXiv",
    primaryClass = "hep-th",
    reportNumber = "FIAN-TD-08-13, ITEP-TH-17-13",
    doi = "10.1063/1.4828688",
    journal = "AIP Conf. Proc.",
    volume = "1562",
    number = "1",
    pages = "123--155",
    year = "2013"
}

@article{DMMSS,
    author = "Dunin-Barkowski, P. and Mironov, A. and Morozov, A. and Sleptsov, A. and Smirnov, A.",
    title = "{Superpolynomials for toric knots from evolution induced by cut-and-join operators}",
    eprint = "1106.4305",
    archivePrefix = "arXiv",
    primaryClass = "hep-th",
    reportNumber = "FIAN-TD-10-11, ITEP-TH-21-11",
    doi = "10.1007/JHEP03(2013)021",
    journal = "JHEP",
    volume = "03",
    pages = "021",
    year = "2013"
}

@article{Bishler2020,
    author = "Bishler, L. and Dhara, Saswati and Grigoryev, T. and Mironov, A. and Morozov, A. and Singh, Vivek Kumar and Ramadevi, P. and Sleptsov, Vivek Kumar Singh A. and Sleptsov, A.",
    title = "{Difference of Mutant Knot Invariants and Their Differential Expansion}",
    eprint = "2004.06598",
    archivePrefix = "arXiv",
    primaryClass = "hep-th",
    reportNumber = "FIAN/TD-10/20; IITP/TH-06/20; ITEP/TH-07/20; MIPT/TH-06/20",
    doi = "10.1134/S0021364020090015",
    journal = "JETP Lett.",
    volume = "111",
    number = "9",
    pages = "494--499",
    year = "2020"
}

@article{Bishler:2020wbq,
    author = "Bishler, L. and Dhara, Saswati and Grigoryev, T. and Mironov, A. and Morozov, A. and Morozov, An. and Ramadevi, P. and Singh, Vivek Kumar and Sleptsov, A.",
    title = "{Distinguishing Mutant knots}",
    eprint = "2007.12532",
    archivePrefix = "arXiv",
    primaryClass = "hep-th",
    reportNumber = "FIAN/TD-12/20; IITP/TH-09/20; ITEP/TH-12/20; MIPT/TH-09/20",
    doi = "10.1016/j.geomphys.2020.103928",
    journal = "J. Geom. Phys.",
    volume = "159",
    pages = "103928",
    year = "2021"
}

@article{MMMRSS,
    author = "Mironov, A. and Morozov, A. and Morozov, An. and Ramadevi, P. and Singh, Vivek Kumar and Sleptsov, A.",
    title = "{Tabulating knot polynomials for arborescent knots}",
    eprint = "1601.04199",
    archivePrefix = "arXiv",
    primaryClass = "hep-th",
    reportNumber = "FIAN-TD-01-16, IITP-TH-01-16, ITEP-TH-01-16, ITEP-TH-02-16",
    doi = "10.1088/1751-8121/aa5574",
    journal = "J. Phys. A",
    volume = "50",
    number = "8",
    pages = "085201",
    year = "2017"
}

@article{Dhara:2017ukv,
    author = "Dhara, Saswati and Mironov, A. and Morozov, A. and Morozov, An. and Ramadevi, P. and Singh, Vivek Kumar and Sleptsov, A.",
    title = "{Eigenvalue hypothesis for multistrand braids}",
    eprint = "1711.10952",
    archivePrefix = "arXiv",
    primaryClass = "hep-th",
    reportNumber = "FIAN-TD-24-17, IITP-TH-18-17, ITEP-TH-31-17",
    doi = "10.1103/PhysRevD.97.126015",
    journal = "Phys. Rev. D",
    volume = "97",
    number = "12",
    pages = "126015",
    year = "2018"
}

@article{Chbili:2022pnt,
    author = "Chbili, Nafaa and Singh, Vivek Kumar",
    title = "{Colored HOMFLY-PT polynomials of quasi-alternating $3$-braid knots}",
    eprint = "2202.09169",
    archivePrefix = "arXiv",
    primaryClass = "hep-th",
    month = "2",
    year = "2022"
}

@article{MMS,
    author = "Mironov, A. and Morozov, A. and Sleptsov, A.",
    title = "{Colored HOMFLY polynomials for the pretzel knots and links}",
    eprint = "1412.8432",
    archivePrefix = "arXiv",
    primaryClass = "hep-th",
    reportNumber = "FIAN-TD-20-14, ITEP-TH-47-14",
    doi = "10.1007/JHEP07(2015)069",
    journal = "JHEP",
    volume = "07",
    pages = "069",
    year = "2015"
}

@article{Itoyama:2012re,
    author = "Itoyama, H. and Mironov, A. and Morozov, A. and Morozov, An.",
    title = "{Eigenvalue hypothesis for Racah matrices and HOMFLY polynomials for 3-strand knots in any symmetric and antisymmetric representations}",
    eprint = "1209.6304",
    archivePrefix = "arXiv",
    primaryClass = "math-ph",
    reportNumber = "FIAN-TD-24-12, ITEP-TH-46-12, OCU-PHYS-375",
    doi = "10.1142/S0217751X13400095",
    journal = "Int. J. Mod. Phys. A",
    volume = "28",
    pages = "1340009",
    year = "2013"
}

@article{Itoyama:2012qt,
    author = "Itoyama, H. and Mironov, A. and Morozov, A. and Morozov, An.",
    title = "{Character expansion for HOMFLY polynomials. III. All 3-Strand braids in the first symmetric representation}",
    eprint = "1204.4785",
    archivePrefix = "arXiv",
    primaryClass = "hep-th",
    reportNumber = "FIAN-TD-05-12, ITEP-TH-15-12, OCU-PHYS-366",
    doi = "10.1142/S0217751X12500996",
    journal = "Int. J. Mod. Phys. A",
    volume = "27",
    pages = "1250099",
    year = "2012"
}

@article{Chen:2015rid,
    author = "Chen, Qingtao and Liu, Kefeng and Zhu, Shengmao",
    title = "{Volume conjecture for $SU(n)$-invariants}",
    eprint = "1511.00658",
    archivePrefix = "arXiv",
    primaryClass = "math.QA",
    month = "11",
    year = "2015"
}

@article{Habiro2007,
	doi = {10.1007/s00222-007-0071-0},
	year = 2007,
	publisher = {Springer Science and Business Media {LLC}},
	volume = {171},
	number = {1}, 
	pages = {1--81},
	author = {Kazuo Habiro},
	title = {A unified Witten-Reshetikhin-Turaev invariant for integral homology spheres},
	journal = {Inventiones mathematicae}
}

@article{Nawata:2015wya,
    author = "Nawata, Satoshi and Oblomkov, Alexei",
    editor = "Gukov, Sergei and Khovanov, Mikhail and Walcher, Johannes",
    title = "{Lectures on knot homology}",
    eprint = "1510.01795",
    archivePrefix = "arXiv",
    primaryClass = "math-ph",
    doi = "10.1090/conm/680/13702",
    journal = "Contemp. Math.",
    volume = "680",
    pages = "137",
    year = "2016"
}

@article{Chen:2014fpa,
    author = "Chen, Qintao and Liu, Kefeng and Peng, Pan and Zhu, Shengmao",
    title = "{Congruent skein relations for colored HOMFLY-PT invariants and colored Jones polynomials}",
    eprint = "1402.3571",
    archivePrefix = "arXiv",
    primaryClass = "math.GT",
    month = "2",
    year = "2014"
}

@article{Kawagoe:2012bt,
    author = "Kawagoe, Kenichi",
    title = "{On the formulae for the colored HOMFLY polynomials}",
    eprint = "1210.7574",
    archivePrefix = "arXiv",
    primaryClass = "math.GT",
    doi = "10.1016/j.geomphys.2016.02.012",
    journal = "J. Geom. Phys.",
    volume = "106",
    pages = "143--154",
    year = "2016"
}

@article{Kawagoe:2021onh,
    author = "Kawagoe, Kenichi",
    title = "{The colored HOMFLY-PT polynomials of the trefoil knot, the figure-eight knot and twist knots}",
    eprint = "2107.08678",
    archivePrefix = "arXiv",
    primaryClass = "math.GT",
    month = "7",
    year = "2021"
}

@article{Dunfield:2005si,
    author = "Dunfield, Nathan M. and Gukov, Sergei and Rasmussen, Jacob",
    title = "{The Superpolynomial for knot homologies}",
    eprint = "math/0505662",
    archivePrefix = "arXiv",
    month = "5",
    year = "2005"
}

@article{Gukov:2011ry,
    author = "Gukov, Sergei and Sto\v{s}i\'c, Marko",
    editor = "Block, Jonathan and Distler, Jacques and Donagi, Ron and Sharpe, Eric",
    title = "{Homological Algebra of Knots and BPS States}",
    eprint = "1112.0030",
    archivePrefix = "arXiv",
    primaryClass = "hep-th",
    reportNumber = "CALT-68-2859",
    doi = "10.1090/pspum/085/1377",
    journal = "Proc. Symp. Pure Math.",
    volume = "85",
    pages = "125--172",
    year = "2012"
}

@article{Gorsky:2013jxa,
    author = "Gorsky, Eugene and Gukov, Sergei and Stosic, Marko",
    title = "{Quadruply-graded colored homology of knots}",
    eprint = "1304.3481",
    archivePrefix = "arXiv",
    primaryClass = "math.QA",
    reportNumber = "CALT-68-2903",
    month = "4",
    year = "2013"
}

@article{Morton,
author = {Morton, Hugh R. and Cromwell, Peter R.},
journal = {Journal of Knot Theory and Its Ramifications},
volume = {05},
number = {02},
pages = {225-238},
year = {1996},
doi = {10.1142/S0218216596000163},
}

@article{berest2021,
  title={Cyclotomic expansion of generalized Jones polynomials},
  author={Berest, Yuri and Gallagher, Joseph and Samuelson, Peter},
  journal={Letters in Mathematical Physics},
  volume={111},
  number={2},
  pages={1--32},
  year={2021},
  publisher={Springer},
  eprint = {1908.04415},
  archivePrefix = {arXiv},
}

@article{lovejoy2019colored,
  title={The colored Jones polynomial and Kontsevich-Zagier series for double twist knots, II},
  author={Lovejoy, Jeremy and Osburn, Robert},
  eprint = {1903.05060},
  archivePrefix = {arXiv},
  year={2019}
}

@article{lovejoy2017colored,
  title={The colored Jones polynomial and Kontsevich-Zagier series for double twist knots},
  author={Lovejoy, Jeremy and Osburn, Robert},
  eprint = {1710.04865},
  archivePrefix = {arXiv},
  year={2017}
}

@article{Chen:2015sol,
    author = "Chen, Qingtao",
    title = "{Cyclotomic expansion and volume conjecture for superpolynomials of colored HOMFLY-PT homology and colored Kauffman homology}",
    eprint = "1512.07906",
    archivePrefix = "arXiv",
    primaryClass = "math.QA",
    month = "12",
    year = "2015"
}

@article{hikami2015torus,
  title={Torus knots and quantum modular forms},
  author={Hikami, Kazuhiro and Lovejoy, Jeremy},
  journal={Research in the Mathematical Sciences},
  volume={2},
  number={1},
  pages={1--15},
  year={2015},
  publisher={Springer},
   eprint = {1409.6243},
    archivePrefix = {arXiv},
}

@article{garoufalidis2005analytic,
  title={An analytic version of the melvin-morton-rozansky conjecture},
  author={Garoufalidis, Stavros and Le, Thang TQ},
  eprint = {math/0503641},
  archivePrefix = {arXiv},
  year={2005}
}

@article{garoufalidis2011asymptotics,
  title={Asymptotics of the colored Jones function of a knot},
  author={Garoufalidis, Stavros and L{\^e}, Thang T Q},
  journal={Geometry \& Topology},
  volume={15},
  number={4},
  pages={2135--2180},
  year={2011},
  publisher={Mathematical Sciences Publishers},
  eprint = {math/0508100},
  archivePrefix = {arXiv},
}

@article{Lanina:2021nfj,
    author = "Lanina, E. and Sleptsov, A. and Tselousov, N.",
    title = "{Chern-Simons perturbative series revisited}",
    eprint = "2105.11565",
    archivePrefix = "arXiv",
    primaryClass = "hep-th",
    doi = "10.1016/j.physletb.2021.136727",
    journal = "Phys. Lett. B",
    volume = "823",
    pages = "136727",
    year = "2021"
}

@article{Lanina:2021jzd,
    author = "Lanina, E. and Sleptsov, A. and Tselousov, N.",
    title = "{Implications for colored HOMFLY polynomials from explicit formulas for group-theoretical structure}",
    eprint = "2111.11751",
    archivePrefix = "arXiv",
    primaryClass = "hep-th",
    reportNumber = "ITEP-TH-31/21, IITP-TH-20/21, MIPT-TH-17/21",
    doi = "10.1016/j.nuclphysb.2021.115644",
    journal = "Nucl. Phys. B",
    volume = "974",
    pages = "115644",
    year = "2022"
}

@inproceedings{morton2009mutant,
  title={Mutant knots with symmetry},
  author={Morton, HR},
  booktitle={Mathematical Proceedings of the Cambridge Philosophical Society},
  volume={146},
  number={1},
  pages={95--107},
  year={2009},
  organization={Cambridge University Press}
}

@article{AMP,
    author = "Anokhina, Aleksandra and Morozov, Alexei and Popolitov, Aleksandr",
    title = "{Nimble evolution for pretzel Khovanov polynomials}",
    eprint = "1904.10277",
    archivePrefix = "arXiv",
    primaryClass = "hep-th",
    doi = "10.1140/epjc/s10052-019-7303-5",
    journal = "Eur. Phys. J. C",
    volume = "79",
    number = "10",
    pages = "867",
    year = "2019"
}

@article{Morozov:2018ges,
    author = "Morozov, A.",
    title = "{Knot polynomials for twist satellites}",
    eprint = "1801.02407",
    archivePrefix = "arXiv",
    primaryClass = "hep-th",
    reportNumber = "ITEP-TH-01-18",
    doi = "10.1016/j.physletb.2018.05.031",
    journal = "Phys. Lett. B",
    volume = "782",
    pages = "104--111",
    year = "2018"
}

@article{Anokhina:2018ysw,
    author = "Anokhina, A. and Morozov, A.",
    title = "{Are Khovanov-Rozansky polynomials consistent with evolution in the space of knots?}",
    eprint = "1802.09383",
    archivePrefix = "arXiv",
    primaryClass = "hep-th",
    reportNumber = "ITEP-TH/05-18, ITEP-TH-05-18",
    doi = "10.1007/JHEP04(2018)066",
    journal = "JHEP",
    volume = "04",
    pages = "066",
    year = "2018"
}

@book{LL3,
editor = {L.D. Landau and E.M. Lifshitz},
title = {Quantum Mechanics},
publisher = {Pergamon},
edition = {Third Edition},
pages = {ii},
year = {1977},
isbn = {978-0-08-020940-1},
doi = {doi.org/10.1016/B978-0-08-020940-1.50001-3}
}

\section{Appendix: Arborescent calculus
\label{arbor}}

Most of calculations in this paper are done for arborescent
and nearly arborescent \cite{MMMRS,Dhara:2017ukv, Chbili:2022pnt} knots
and the entire consideration is so far restricted to symmetric representations.
We remind here just a few basic things about the powerful arborescent calculus
\cite{MMMRS}, where just two exclusive Racah matrices $S$ and $\bar S$
are enough.
There is  no  restriction to symmetric representations:
see \cite{Morozov4,Morozov3} for what is currently known about $S$ and $\bar S$ for
generic rectangular and non-rectangular reps.
Still we collect the formulas for this particular case --
they are known exhaustively and are much simpler than the general ones.

\subsection{$S$ and $\bar S$ in all symmetric representations}

The ${\cal R}$ matrix eigenvalues in symmetric representations $R$
in the two channels $R\otimes R$ and $R\otimes \bar R$ are very simple:
\be
T = {\rm diag}\Big((-)^{r + i - 1} q^{-r^2 + (i-1)^2 + i-1} A^{-r} \Big)   \ \ \ \ \ \
\bar T = {\rm diag}\Big((-)^{r+i-1}q^{(i-1)(i-2)}A^{i-1}\Big),
\ee
For symmetric representations the matrices $S$,
which switch between $R\otimes \bar R$ and $\bar R \otimes \bar R$
and $\bar S$ acting within the space $\bar R \otimes \bar R$,
are also well known
\cite{MMS,LL3}:
for $i,j=1,\ldots, r+1$
\be
S_{ij} = \sqrt{\frac{\bar  d_i}{ d_j}}\cdot  \alpha_{i-1,j-1} \ \ \ \ \ \ \
\bar S_{ij} = \sqrt{\frac{\bar d_i}{\bar d_j}}\cdot \bar\alpha_{i-1,j-1}
\ee
where $d_X$ and $\bar  d_X$ are dimensions of representations from $R\otimes  R$
and $\bar R\times  R$ respectively,
\be
d_j = {\rm dim}_{[r+j-1,r-j+1]} :=\frac{[2j-1]}{\prod_{i=1}^{r+j}{[i]}\prod_{i=1}^{r+1-j} [i]}
\cdot \prod_{i=1}^{r+j-2}\{Aq^i\}\prod_{i=-1}^{r-1-j}\{Aq^i\}  \nn \\
\bar d_i :=\{Aq^{2i-3}\}\{Aq^{-1}\}\prod_{j=0}^{i-3} \left(\frac{\{Aq^j\}}{[j+2]}\right)^2
\ee
and
\be
\left(\begin{array}{c}
\alpha_{km}(r) \\ \\ \bar\alpha_{km}(r)
\end{array}\right)
= \frac{(-)^{r+k+m} [2m+1]\Big([k]![m]!\Big)^2[r-k]![r-m]!}{[r+k+1]![r+m+1]!}
\cdot \ \ \ \ \ \ \ \ \ \ \ \ \ \ \ \ \ \ \ \ \ \    \nn \\ \cdot
\sum_{j={\rm max}(r+m,r+k)}^{{\rm min}(r+k+m,2r)} \frac{(-)^j[j+1]!}{[2r-j]!\Big([j-r-k]! [j-r-m]![r+k+m-j]!\Big)^2}
\cdot\left(\begin{array}{c}
\frac{{\cal D}_{r-m}{\cal D}_{j+1}}{{\cal D}_{r+k+1}{\cal D}_{j-r-m} } \\  \\
\frac{{\cal D}_m^2{\cal D}_{j+1}}{{\cal D}_{r+k+1}{\cal D}_{r+m+1}{\cal D}_{r+k+m-j}}
\end{array}\right)
\label{alphasym}
\ee
where ${\cal D}_n:= \frac{1}{[n]!}\prod_{j=-1}^{n-2}\{Aq^j\}$
are responsible for the deviation from the $sl_2$ case when $A=q^2$ and ${\cal D}_n=1$.

\subsection{Pretzel fingers
\label{prefing}}

In this section, following \cite{MMMRS},
we list the simplest possible types of propagators and fingers,
belonging to the  {\it pretzel} type.
These are the formulas used in the examples in the main text.

In the case of pretzels, the notation with bars is sufficient to
distinguish between all these cases,
still we add explicit indices $par,ea,oa$ to avoid any confusion:
\be
\begin{picture}(350,165)(-300,-55)
\put(-270,0){\line(1,0){30}}
\put(-270,0){\line(0,1){30}}
\put(-240,30){\line(-1,0){30}}
\put(-240,30){\line(0,-1){30}}
\put(-257,13){\mbox{$n$}}
\put(-273,-40){\line(1,5){8}}
\put(-237,-40){\line(-1,5){8}}
\put(-290,-50){\mbox{$X,a $}}
\put(-240,-50){\mbox{$\bar X,b $}}
\put(-272,-35){\vector(0,1){2}}
\put(-238,-36){\vector(0,1){2}}
\put(-290,75){\mbox{$Z,e $}}
\put(-240,75){\mbox{$\bar Z,f $}}
\put(-237,70){\line(-1,-5){8}}
\put(-273,70){\line(1,-5){8}}
\put(-272,66){\vector(0,1){2}}
\put(-238,65){\vector(0,1){2}}
\put(-285,70){\vector(0,-1){110}}
\put(-225,70){\vector(0,-1){110}}
\put(-285,66){\vector(0,-1){2}}
\put(-225,66){\vector(0,-1){2}}
\qbezier(-295,-20)(-255,0)(-215,-20)
\qbezier(-295,-20)(-255,-40)(-215,-20)
\qbezier(-295,50)(-255,70)(-215,50)
\qbezier[50](-295,50)(-255,30)(-215,50)
\put(-295,-20){\line(0,1){70}}
\put(-215,-20){\line(0,1){70}}
\put(-360,70){\mbox{$Z\ \in\ R\otimes \bar R$}}
\put(-360,13){\mbox{$Y\ \in\ R\otimes R$}}
\put(-360,-40){\mbox{$X\ \in\ R\otimes \bar R$}}
%
%\put(-200,43){\mbox{$=\sum_{Y,c,d} S^{RR\bR\bR}_{\bar Zfe,Ycd}\, t_{Yc}^n\,S^{\bar RR R\bar R}_{Ycd,Xab}$}}
%\put(-180,7){\mbox{$  \stackrel{(\ref{Stypes})}{=}
%\sum_{Y,c,d} S_{Zef,Ycd}\, t_{Yc}^n\, S_{\bar Xba,Ycd}$}}
\put(-170, 0){\mbox{$\boxed{{\cal A}^{\rm par}(n) = S  T^n S^\dagger}$}}
\put(-160,95){\mbox{{\bf parallel braid:}}}
\put(0,95){\mbox{$\downarrow\uparrow \ \ \uparrow\downarrow$}}
\put(0,0){\line(1,0){30}}
\put(0,0){\line(0,1){30}}
\put(30,30){\line(-1,0){30}}
\put(30,30){\line(0,-1){30}}
\put(13,13){\mbox{$n$}}
\put(-3,-40){\line(1,5){8}}
\put(33,-40){\line(-1,5){8}}
\put(-20,-50){\mbox{$X,a$}}
\put(30,-50){\mbox{$\bar X,b$}}
\put(-2,-35){\vector(0,1){2}}
\put(32,-36){\vector(0,1){2}}
\qbezier(5,30)(5,50)(-5,50)
\qbezier(-15,30)(-15,50)(-5,50)
\qbezier(25,30)(25,50)(35,50)
\qbezier(45,30)(45,50)(35,50)
\put(-15,30){\vector(0,-1){70}}
\put(45,30){\vector(0,-1){70}}
\qbezier(-25,-20)(15,0)(55,-20)
\qbezier(-25,-20)(15,-40)(55,-20)
\qbezier(-25,-20)(-35,70)(15,70)
\qbezier(55,-20)(65,70)(15,70)
\put(70,13){\mbox{$=\ (S T^nS^\dagger)_{1X}^{ab} $}}
\end{picture}
\label{parpre}
\ee

\be
\begin{picture}(350,165)(-300,-55)
\put(-270,0){\line(1,0){30}}
\put(-270,0){\line(0,1){30}}
\put(-240,30){\line(-1,0){30}}
\put(-240,30){\line(0,-1){30}}
\put(-269,13){\mbox{{\rm even}\ $n$}}
\put(-273,-40){\line(1,5){8}}
\put(-237,-40){\line(-1,5){8}}
\put(-290,-50){\mbox{$X,a$}}
\put(-240,-50){\mbox{$\bar X,b$}}
\put(-272,-35){\vector(0,1){2}}
\put(-238,-36){\vector(0,-1){2}}
\put(-290,75){\mbox{$Z,e$}}
\put(-240,75){\mbox{$\bar Z,f$}}
\put(-237,70){\line(-1,-5){8}}
\put(-273,70){\line(1,-5){8}}
\put(-272,66){\vector(0,1){2}}
\put(-238,65){\vector(0,-1){2}}
\put(-285,70){\vector(0,-1){110}}
\put(-225,-40){\vector(0,1){110}}
\put(-285,66){\vector(0,-1){2}}
\put(-225,-36){\vector(0,1){2}}
\qbezier(-295,-20)(-255,0)(-215,-20)
\qbezier(-295,-20)(-255,-40)(-215,-20)
\qbezier(-295,50)(-255,70)(-215,50)
\qbezier[50](-295,50)(-255,30)(-215,50)
\put(-295,-20){\line(0,1){70}}
\put(-215,-20){\line(0,1){70}}
\put(-360,70){\mbox{$Z\ \in\ R\otimes \bar R$}}
\put(-360,13){\mbox{$Y\ \in\ R\otimes \bR$}}
\put(-360,-40){\mbox{$X\ \in\ R\otimes \bar R$}}
%
%\put(-200,43){\mbox{$=\sum_{Y,c,d} S^{R\bR R\bR}_{\bar Zfe,Ycd}\, \bar t_{Yc}^n\,S^{\bar RR \bR R}_{Ycd,Xab}$}}
%\put(-180,7){\mbox{$  \stackrel{(\ref{Stypes})}{=}
%\sum_{Y,c,d} \bar S_{Zef,Ycd}\, \bar t_{Yc}^n \, \bar S_{\bar X_{ab},\bar Ydc}$}}
\put(-170,0){\mbox{$\boxed{{\cal A}^{\rm ea}(\bar n) =  \bar S  \bar T^n \bar S}$}}
\put(-160,95){\mbox{{\bf even antiparallel braid:}}}
\put(0,95){\mbox{$\downarrow\uparrow \ \ \downarrow\uparrow$}}
\put(0,0){\line(1,0){30}}
\put(0,0){\line(0,1){30}}
\put(30,30){\line(-1,0){30}}
\put(30,30){\line(0,-1){30}}
\put(1,13){\mbox{{\rm even} $n$}}
\put(-3,-40){\line(1,5){8}}
\put(33,-40){\line(-1,5){8}}
\put(-20,-50){\mbox{$X,a$}}
\put(30,-50){\mbox{$\bar X,b$}}
\put(-2,-35){\vector(0,1){2}}
\put(32,-36){\vector(0,-1){2}}
\qbezier(5,30)(5,50)(-5,50)
\qbezier(-15,30)(-15,50)(-5,50)
\qbezier(25,30)(25,50)(35,50)
\qbezier(45,30)(45,50)(35,50)
\put(-15,30){\vector(0,-1){70}}
\put(45,30){\line(0,-1){70}}
\put(45,-36){\vector(0,1){2}}
\qbezier(-25,-20)(15,0)(55,-20)
\qbezier(-25,-20)(15,-40)(55,-20)
\qbezier(-25,-20)(-35,70)(15,70)
\qbezier(55,-20)(65,70)(15,70)
\put(70,13){\mbox{$=\ (\bar S \bar T^n\bar S)_{1X}^{ab} $}}
\end{picture}
\label{{antiparevenpre}}
\ee

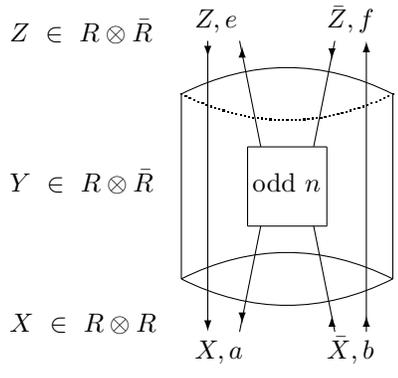
\be
\begin{picture}(350,165)(-300,-55)
\put(-270,0){\line(1,0){30}}
\put(-270,0){\line(0,1){30}}
\put(-240,30){\line(-1,0){30}}
\put(-240,30){\line(0,-1){30}}
\put(-268,13){\mbox{{\rm odd} $n$}}
\put(-273,-40){\line(1,5){8}}
\put(-237,-40){\line(-1,5){8}}
\put(-290,-50){\mbox{$X,a$}}
\put(-240,-50){\mbox{$\bar X,b$}}
\put(-272,-35){\vector(0,-1){2}}
\put(-238,-36){\vector(0,1){2}}
\put(-290,75){\mbox{$Z,e$}}
\put(-240,75){\mbox{$\bar Z,f$}}
\put(-237,70){\line(-1,-5){8}}
\put(-273,70){\line(1,-5){8}}
\put(-272,66){\vector(0,1){2}}
\put(-238,65){\vector(0,-1){2}}
\put(-285,70){\vector(0,-1){110}}
\put(-225,-40){\vector(0,1){110}}
\put(-285,66){\vector(0,-1){2}}
\put(-225,-36){\vector(0,1){2}}
\qbezier(-295,-20)(-255,0)(-215,-20)
\qbezier(-295,-20)(-255,-40)(-215,-20)
\qbezier(-295,50)(-255,70)(-215,50)
\qbezier[50](-295,50)(-255,30)(-215,50)
\put(-295,-20){\line(0,1){70}}
\put(-215,-20){\line(0,1){70}}
\put(-360,70){\mbox{$Z\ \in\ R\otimes \bar R$}}
\put(-360,13){\mbox{$Y\ \in\ R\otimes \bR$}}
\put(-360,-40){\mbox{$X\ \in\ R\otimes R$}}
%
%\put(-200,43){\mbox{$=\sum_{Y,c,d} S^{R\bR R\bR}_{\bar Zfe,Ycd}\, \bar t_{Yc}^n\,S^{RR \bR\bar R}_{Ycd,Xab}$}}
%\put(-180,7){\mbox{$  \stackrel{(\ref{Stypes})}{=}
%\bar S_{Zef,Ycd}\, \bar t_{Yc}^n\, S_{Ycd,Xab}$}}
\put(-170,0){\mbox{$\boxed{{\cal A}^{\rm oa}(n) =  \bar S  \bar T^n S}$}}
\put(-160,95){\mbox{{\bf odd antiparallel braid:}}}
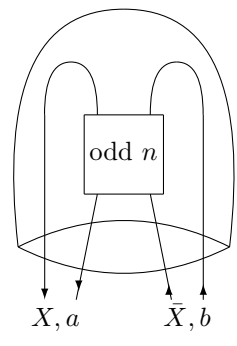
\put(0,95){\mbox{$\downarrow\downarrow \ \ \uparrow\uparrow$}}
\put(0,0){\line(1,0){30}}
\put(0,0){\line(0,1){30}}
\put(30,30){\line(-1,0){30}}
\put(30,30){\line(0,-1){30}}
\put(2,13){\mbox{{\rm odd} $n$}}
\put(-3,-40){\line(1,5){8}}
\put(33,-40){\line(-1,5){8}}
\put(-20,-50){\mbox{$X,a$}}
\put(30,-50){\mbox{$\bar X,b$}}
\put(-2,-35){\vector(0,-1){2}}
\put(32,-36){\vector(0,1){2}}
\qbezier(5,30)(5,50)(-5,50)
\qbezier(-15,30)(-15,50)(-5,50)
\qbezier(25,30)(25,50)(35,50)
\qbezier(45,30)(45,50)(35,50)
\put(-15,30){\vector(0,-1){70}}
\put(45,30){\line(0,-1){70}}
\put(45,-36){\vector(0,1){2}}
\qbezier(-25,-20)(15,0)(55,-20)
\qbezier(-25,-20)(15,-40)(55,-20)
\qbezier(-25,-20)(-35,70)(15,70)
\qbezier(55,-20)(65,70)(15,70)
\put(70,13){\mbox{$=\ (\bar S \bar T_0^n S)_{1X}^{ab} $}}
\end{picture}
\label{{antiparoddpre}}
\ee

\bigskip

\end{document}